\documentclass[conference]{IEEEtran}

\usepackage{alltt                                    % I like these
          , multirow
          , booktabs
          , listings
          , graphicx
          ,float
	,cite
          ,verbatim
         ,mathtools
	,url
	,amsmath
}
\usepackage[table]{xcolor}
\usepackage[numbers]{natbib}     % this is a better citation system
\usepackage{syntax}
\usepackage{enumitem}
\usepackage{threeparttable}
\usepackage{framed}
\usepackage{hhline}
\usepackage[ruled]{algorithm}
\usepackage{algpseudocode}
\usepackage{balance}

\usepackage{tikz}
\usepackage{verbatim}
\usetikzlibrary{arrows,shapes,backgrounds}

\tikzstyle{vertex}=[ellipse,fill=black!25,minimum size=20pt, inner sep=0pt]
\tikzstyle{edge} = [draw,thin,-]
\tikzstyle{glabel} = [text width=1cm,text centered,font=\bf]
\pgfdeclarelayer{bg}    % declare background layer
\pgfsetlayers{bg,main}  % set the order of the layers (main is the standard layer)

\usepackage{expl3}
\ExplSyntaxOn
\newcommand\latinabbrev[1]{
  \peek_meaning:NTF . {% Same as \@ifnextchar
    #1\@}%
  { \peek_catcode:NTF a {% Check whether next char has same catcode as \'a, i.e., is a letter
      #1., \@ }%
    {#1., \@}}}
\ExplSyntaxOff

\newcommand{\CASE}[1]{\STATE \textbf{case} #1\textbf{:} \begin{ALC@g}}
\newcommand{\ENDCASE}{\end{ALC@g}}

\newcommand{\DEFAULT}{\STATE \textbf{default:} \begin{ALC@g}}
\newcommand{\ENDDEFAULT}{\end{ALC@g}}
\newcommand{\DEFAULTLINE}[1]{\STATE \textbf{default:} }
\let\footnotesize\scriptsize

\newsavebox{\supbox}% Superscript box
\newcommand{\bsup}{\begin{lrbox}{\supbox}$\tt\scriptstyle}% Superscript begin
\newcommand{\esup}{$\end{lrbox}{}^{\usebox{\supbox}}}% Superscript end
\def\eg{\latinabbrev{e.g}}
\def\ie{\latinabbrev{i.e}}

\algnewcommand{\LineComment}[1]{\State \(\triangleright\) #1}
\algdef{SE}[DOWHILE]{Do}{doWhile}{\algorithmicdo}[1]{\algorithmicwhile\ #1}%

\definecolor{lightpurple}{rgb}{0.8,0.8,1}
\definecolor{codebg}{RGB}{255,255,255}
\definecolor{commentcolor}{RGB}{11,140,11}
\begin{document}
%
% --- Author Metadata here ---
%\conferenceinfo{WeTSOM}{'2014, Hyderabad India}
%\CopyrightYear{2007} % Allows default copyright year (20XX) to be over-ridden - IF NEED BE.
%\crdata{0-12345-67-8/90/01}  % Allows default copyright data (0-89791-88-6/97/05) to be over-ridden - IF NEED BE.
% --- End of Author Metadata ---

\title{Improved Query Reformulation for Concept Location using CodeRank and Document Structures \vspace{-.4cm}}

\author{\IEEEauthorblockN{Mohammad Masudur Rahman  ~~~ Chanchal K. Roy}
\IEEEauthorblockA{Department of Computer Science, University of Saskatchewan, Canada\\
\{masud.rahman, chanchal.roy\}@usask.ca}
}

\maketitle

\begin{abstract}
During software maintenance, developers usually deal with a significant number of software change requests. 
As a part of this, they often formulate an initial query from the request texts, and then attempt to map the concepts discussed in the request to relevant source code locations in the software system (a.k.a., concept location). 
Unfortunately, studies suggest that they often perform poorly in choosing the right search terms for a change task. 
In this paper, we propose a novel technique --ACER-- that takes an initial query, identifies appropriate search terms from the source code using a novel term weight --CodeRank,
and then suggests effective reformulation to the initial query by exploiting the source document structures, query quality analysis and machine learning.
Experiments with 1,675 baseline queries from eight subject systems report that our technique can improve 71\% of the baseline queries which is highly promising.
Comparison with five closely related existing techniques in query reformulation not only validates our empirical findings but also demonstrates the superiority of our technique.

\end{abstract}

%A category including the fourth, optional field follows...
%\category{D.2.8}{Software Engineering}{Metrics}[maintainability measures, reusability measures]
%\terms{Theory, Metrics, Human factors}
%\keywords{Code quality, readability, reusability, maintainability}
\IEEEpeerreviewmaketitle

\section{Introduction}
Studies show that about 80\% of the total efforts is spent in software maintenance \cite{seahawk} where developers deal with a significant number of software issues \cite{saha,parnindebug,debugvess}. 
%During maintenance, software developers deal with a number of change requests that either focus on software bugs and errors or lead to new software features.
Software issue reports (a.k.a., change requests) discuss both unexpected (or erroneous features such as bugs) and expected but non-existent features (\eg\ new functionality).  
For both bug resolution and new feature implementation, a developer is required to map the concepts discussed in the issue report to appropriate source code within the project which is widely known as concept location \cite{kevicdict,irmarcus,ase2016masud}.
% for implementing the change 
%Such mapping is possibly trivial for a developer who has substantial knowledge on the target project. However, the developers involved in maintenance might not be aware of the low-level architectures of the project, and modern projects could be inherently complex.
%Thus, the developers often experience difficulties in identifying the source locations (\ie\ classes, methods) that need to be changed. 
Developers generally choose one or more important keywords from the report texts, and then use a search method (\eg\ regular expression) to locate the source code entities (\eg\ classes, methods) that need to be changed.
Unfortunately, as the existing studies \cite{kevic,sitir} report, developers regardless of their experience perform poorly in choosing appropriate search terms for software change tasks. 
According to \citet{kevic}, only 12.20\% of the search terms chosen by the developers were able to locate relevant source code entities for the change tasks. 
\citet{vocaprob} also suggest that there is a little chance (\ie\ 10\%--15\%) that developers guess the exact words used in the source code. 
One way to assist the developers in this regard is to automatically suggest helpful reformulations (\eg\ complementary keywords) to their initially chosen queries.

Existing studies apply relevance feedback from developers \cite{gayg}, pseudo-relevance feedback from information retrieval methods \cite{refoqus}, and machine learning  \cite{refoqus,trconfig} for such query reformulation tasks. They also make use of context of query terms from source code \cite{ccmapping,infer,ase2016masud,sisman,hillicse09}, text retrieval configuration \cite{refoqus,trconfig}, and quality of queries \cite{specificity,qperf} in suggesting the reformulated queries.  
\citet{gayg} capture explicit feedback on document relevance from the developers, and then suggest reformulated queries using Rocchio's expansion \cite{rocchio}.
\citeauthor{refoqus} and colleagues \cite{refoqus,qeffect,qperf,soniatool,specificity} take quality of a given query (\ie\ query difficulty) into consideration, and suggest the best reformulation strategy for the query using machine learning.
%\citet{ccmapping} analyze signature and associated comments of a method, mine semantically similar word pairs, and  
%from the source code for mining semantically similar word pairs, and 
%then reformulate a given query.
While all these above techniques are reported to be novel or effective, most of them also share several limitations. 
First, source documents contain both structured items (\eg\ method signatures, formal parameters) and unstructured items (\eg\ code comments).
Unfortunately, many of the above reformulation approaches \cite{refoqus,sisman,gayg} treat the source documents as simple plain text documents, and ignore most of their structural aspects except structured tokens.
Such inappropriate treatment might lead to suboptimal or poor queries.
In fact, \citet{hillicse09} first consider document structures, and suggest natural language phrases from method signatures and field signatures for local code search.
However, since they apply only simple textual matching between initial queries and the signatures, 
the suggested phrases are subject to the quality of not only the given queries and but also of the identifier names from those signatures.            
%in query suggestion for code search. 
%In fact, \citet{hillicse09} first consider structures of a source document where they extract natural language phrases from method and field signatures.
%most of these studies simply split the structured items and treat the documen 
%During collecting terms for query reformulation from the project source, they (1) consider each source file as a regular text file by ignoring its structural aspects (\eg\ programming constructs, methods, classes), and (2) 
Second, many of these approaches often directly apply traditional metrics of term importance (\eg\ avgIDF \cite{specificity}, TF-IDF \cite{rocchio}) to source code 
which were originally targeted for unstructured regular texts (\eg\ news article) \cite{tfidf}.  
Thus, they might also fail to identify the appropriate terms from the structured source documents for query reformulation.

%suggest semantically similar words for a given query by analyzing 
%First, collecting explicit feedback from the developers could be highly expensive, and such study \cite{relevancefb} could also be hard to replicate. 
%Second, machine learning model of \citeauthor{refoqus} is reported to be performing well in the case of \emph{within-project} training, and only 51--72 queries are considered from each of the five projects for training and testing \cite{refoqus}. 
%%It is not quite certain that the model would provide similar performance with larger projects given that the model is not tested with larger systems.
%Given such small dataset, the reported performance possibly could not be generalized for large systems. 
%%against limited amount of data, \ie\ only 50 queries from each subject system.
%Third, \citeauthor{ccmapping} require the source code to be well documented for the mining of word pairs, and hence, might not perform well if the code is poorly documented \cite{ccmapping}.
%%Third, \citeauthor{refoqus} also need sufficient training and testing data. 
%Thus, we need a technique that is neither subject to the training data nor the availability of comments in the source code.
%One way to possibly overcome those concerns is to apply \emph{crowd generated knowledge} in the reformulation of queries for concept location.

In this paper, we propose a novel technique--ACER--for automatic query reformulation for concept location in the context of software change tasks. 
We (1) first introduce a novel graph-based term weight --\emph{CodeRank}-- for identifying important terms from the source code, and then (2) 
apply that term weight and source document structures (\eg\ method signatures) to our technique for automatic query reformulation.
%demonstrate its high potential for the query reformulation using our technique. 
\emph{CodeRank} identifies important terms not only by analyzing salient structured entities (\eg\ camel case tokens), 
but also by exploiting the co-occurrences among the terms across various entities. 
%We adapt our technique for source code from TextRank \cite{rada,blanco}, another term weighting technique for regular texts, and then incorporate our term weights into our query expansion algorithm.
Our technique--ACER--accepts a natural language query as input, 
develops multiple candidate queries from
%extracts important and complementary terms 
two different important contexts, (1) method signatures and (2) field signatures of the 
source documents independently using CodeRank, and then suggests the best reformulation ( based on query quality analysis and machine learning \cite{qperf,refoqus}) to the poorly performing initial query.
%Unlike earlier approaches, our technique (1) adopts a more appropriate term weight for the source code, and (2)  
%our term weight leverages co-occurrences among the terms and our reformulation technique 
%exploits the structures of source documents in addition to their textual content. 
%Thus, our technique has a greater potential for identifying the important terms from the source code and for suggesting effective reformulations to the queries.

%\noindent
%{(-0.5,2.5)/part},
%{(2.1,1)/brought},
%{(3.3,1)/to},
%{(4.3,1)/top},
%{(1,2)/visible},
%{(2,2)/input},
%{(3.1,2)/changed},
%{(4.4,2)/hidden},
%{(1,3)/toggle},
%{(2,3)/when},
%{(3.5,3)/deactivated},
%{(4.8,3)/run},
%{(1,4)/activated},
%{(2.1,4)/closed},
%{(3.1,4)/folding},
%{(4.1,4)/next},
%{(1,1)/opened}}
%part/brought,
%brought/part,
%brought/to,
%to/brought,
%to/top,
%top/to,
%part/visible,
%visible/part,
%part/input,
%input/part,
%input/changed,
%changed/input,
%part/hidden,
%hidden/part,
%toggle/folding,
%folding/toggle,
%part/deactivated,
%deactivated/part,
%part/closed,
%closed/part,
%part/activated,
%activated/part,
%run/when,
%when/run,
%when/next,
%next/when,
%next/visible,
%visible/next,
%part/opened,
%opened/part

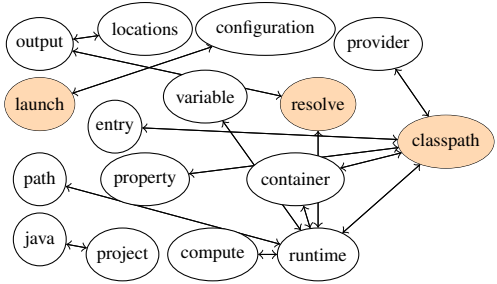
\begin{figure}
\centering
%\resizebox{3.5in}{!}{%
\begin{tikzpicture}[scale=1, auto,swap]
    % Draw a 7,11 network
    % First we draw the vertices
 \foreach \pos/\name in {
{(.6,1.2)/java},
{(1.7,1)/project},
{(2.9,1)/compute},
{(4.3,1)/runtime},
{(.6,2)/path},
{(6,2.5)/classpath},
{(2,2)/property},
{(4,2)/container},
{(.6,3)/launch},
{(1.6,2.7)/entry},
{(2.8,3.1)/variable},
{(4.3,3)/resolve},
{(.6,3.8)/output},
{(2,4)/locations},
{(3.6,4)/configuration},
{(5.1,3.8)/provider}}
\node[vertex] (\name) at \pos [draw=black, fill=white] {\scriptsize\name};
 
\node[vertex] at (6,2.5) [fill=orange!30] {\scriptsize classpath};
\node[vertex] at (4.3,3) [fill=orange!30] {\scriptsize resolve};
\node[vertex] at (.6,3) [fill=orange!30] {\scriptsize launch};

\begin{pgfonlayer}{bg}    % select the background layer
 %\draw (foo) -- (baz);
\foreach \source/ \dest  in {
launch/configuration,
configuration/launch,
java/project,
project/java,
compute/runtime,
runtime/compute,
runtime/path,
path/runtime,
classpath/property,
property/classpath,
runtime/container,
container/runtime,
container/classpath,
classpath/container,
classpath/entry,
entry/classpath,
variable/runtime,
runtime/variable,
runtime/classpath,
classpath/runtime,
resolve/runtime,
runtime/resolve,
resolve/output,
output/resolve,
output/locations,
locations/output,
classpath/provider,
provider/classpath}
\path[edge, ->, draw=black] (\source) -- (\dest);
\end{pgfonlayer}

\end{tikzpicture}
%}
\vspace{-.2cm}
\caption{An example term graph generated by CodeRank for source code of Listing \ref{lst:example}}
\label{fig:tgraph}
\vspace{-.6cm}
\end{figure}

Table \ref{table:ctask} shows an example change request \cite{changereq} submitted for \texttt{eclipse.jdt.debug} system, and it refers to ``debugger source lookup"  issue of Eclipse IDE.  Let us assume that the developer chooses important keywords from the request title, and 
formulates a generic initial query--``debugger source lookup."
%--by choosing important keywords from the request title.
Unfortunately, the query does not perform well, and returns the first correct result at the 79$^{th}$ position of the result list. 
Further extension--``debugger source lookup work variables"--also does not help, and returns the result at the 77$^{th}$ position.
The existing technique -- RSV \cite{qsurvey}-- extends the query as follows--``debugger source lookup work variables \emph{launch configuration jdt java debug"}--where the new terms are collected from the project source using TF-IDF based term weight.
This query returns the correct result at the 30$^{th}$ position which is also far from ideal unfortunately.
The query of \citet{sisman}--``debugger source lookup work variables \emph{test exception suite core code}"--also returns the correct result at the 51$^{st}$ position.
%The reformulated query of \citet{hillicse09} based on partial phrasal matching--``preferences disable enable folding small file \emph{toggle folding''}--returns the correct result at 18$^{th}$ position. 
On the other hand, our suggested query--``debugger source lookup work variables \emph{launch debug problem resolve required classpath"}--returns the correct result at the 2$^{nd}$ position which is highly promising.
We first collect structured tokens (\eg\ \texttt{resolveRuntimeClasspathEntry}) from method signatures and field signatures of the source code (\eg\ Listing \ref{lst:example}), and split them into simpler terms (\eg\ \texttt{resolve, Runtime, Classpath} and \texttt{Entry}). 
The underlying idea is that such signatures often encode high level intents and important domain terms while the rest of the code focuses on more granular level implementation details, and thus possibly contains more noise \cite{hillicse09,shepherd}.
%For instance, the example code in Listing \ref{lst:example} contains several of our query terms (\ie\ \emph{resolve, Classpath}) in the method signatures.  
We develop individual term graph (\eg\ Fig. \ref{fig:tgraph}) based on term co-occurrences  from each signature type, apply CodeRank term weighting, and extract multiple candidate reformulations with the highly weighted terms (\eg\ orange coloured, Fig. \ref{fig:tgraph}).
Then we analyze the quality of the candidates using their quality measures \cite{qperf}, apply machine learning, and suggest the best reformulation to the initial query.   
%Then we develop a source term graph (\eg\ Fig. \ref{fig:tgraph}) based on co-occurrences of the simpler terms across multiple entities (\eg\ method signatures).
%Parts of the relevant source code and term graph are displayed in Listing \ref{lst:example} and Fig. \ref{fig:tgraph} respectively.
%We then apply \emph{CodeRank}, our adaptation of a popular graph-based term weighting algorithm \cite{pagerank,rada} to the term graph,
%and suggest the highly weighted terms (\ie\ orange coloured, Fig. \ref{fig:tgraph}) for query expansion.  
Thus, our technique (1) first captures salient terms from the source documents by analyzing their structural aspects (\ie\ unlike \emph{bag of words} approaches \cite{salton}) and an appropriate term weight--CodeRank, and 
(2) then suggests the best query reformulation using document structures (\ie\ multiple candidates derived from various signatures), query quality analysis and machine learning \cite{qperf}. 

\begin{table}[!t]
\centering
\caption{An Example Change Request (Issue \#:31110, eclipse.jdt.debug)}\label{table:ctask}
\vspace{-.2cm}
\resizebox{3.5in}{!}{%
\begin{threeparttable}
\begin{tabular}{l| p{6cm}}
\hline
\textbf{Field} & \textbf{Content}\\
\hline
%\hline
%Issue ID & 303705\\
%\hline
%Product &  eclipse.jdt.ui \\
%\hline
%Component & UI\\
\hline
Title & Debbugger Source Lookup does not work with variables\\
%& on small file takes forever\\
\hline
Description & In the Debugger Source Lookup dialog I can also select\\ 
& variables for source lookup. (Advanced... $>$ Add Variables). \\
& I selected the variable which points to the archive containing \\
& the source file for the type, but the debugger still claims that \\
& he cannot find the source.\\
\hline
\hline
\textbf{Initial} & \multirow{2}{*}{debugger source lookup work variables}\\
\textbf{Search Query}&  \\
\hline
\end{tabular}
%\begin{tablenotes}
%\item [1] No. of example pairs for which relative quality evaluation matches with that of StackOverflow
%\item [2] \% of agreement, \item [3] \% of disagreement
\centering
%\textbf{CC}=Review comments with code elements, \textbf{CWC}=Review comments without code elements 
 %\end{tablenotes}
\end{threeparttable}
}
\vspace{-.2cm}
\end{table}

\begin{lstlisting}[label=lst:example, language=java,  escapechar=@, aboveskip=5pt, float=t, belowskip=-2em, frame=bt, caption={Source code used for automatic query reformulation (abridged from \cite{examplecode})}]
public static IRuntime@\textbf{Classpath}@Entry[] @\textbf{resolve}@Runtime@\textbf{Classpath}@Entry(IRuntime@\textbf{Classpath}@Entry entry, IJavaProject project) throws CoreException {
		switch (entry.getType()) {
			case IRuntimeClasspathEntry.PROJECT:
				// if the project has multiple output locations, they must be returned
				IResource resource = entry.getResource();
				if (resource instanceof IProject) {
					IJavaProject jp = JavaCore.create((IProject)resource);
					if (jp.exists() && jp.getProject().isOpen()) {
						IRuntimeClasspathEntry[] entries = resolveOutputLocations(jp);
					}
				}
			break; 
		---------------------------------------------------
}}
\end{lstlisting}

Experiments using 1,675 baseline queries from eight open source subject systems show that our technique can improve 71\%  (and preserve 26\%) of the baseline queries which are highly promising according to relevant literature \cite{refoqus,trconfig,qsurvey}.
Our suggested queries return correct results for 64\% of the queries in the Top-100 results.  
Our findings report that \emph{CodeRank} is a more effective term weighting method than the traditional methods (\eg\ TF, TF-IDF) for search query reformulation in the context of source code. Our findings also suggest that structure of a source code document is an important paradigm for both term weighting and query reformulation.
Comparison with five closely related existing approaches \cite{hillicse09,rocchio,qsurvey,sisman,refoqus} not only validates our empirical findings but also demonstrates the superiority of our technique. 
Thus, the paper makes the following contributions:
\begin{itemize}
\item A novel term weighting method --CodeRank-- for source code that identifies the most important terms from a given code entity (\eg\ class, method). 
\item A novel query reformulation technique that reformulates a given initial query using CodeRank, source document structures, query quality analysis and machine learning.
\item Comprehensive evaluation using 1,675 baseline queries from eight open source subject systems. 
\item Comparison with five closely related existing approaches from the literature.
\end{itemize}

%The rest of the paper is organized as follows -- 
%Section \ref{sec:bg} provides background overview,
%Section \ref{sec:acer} explains our adopted methodology and algorithms, 
%and Section \ref{sec:experiment} discusses the conducted experiments, findings and validations.
%Section \ref{sec:threats} identifies the threats to our findings, Section \ref{sec:related} describes the related work, 
%and finally Section \ref{sec:conclusion} concludes the paper with future work.

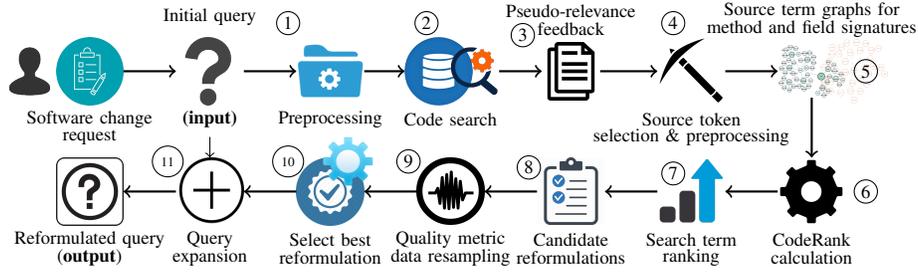
\begin{figure*}
\centering
%\resizebox{3.5in}{!}{%
\begin{tikzpicture}[scale=.8, auto,swap]

\node at (2.3,2.8) [circle, inner sep=1pt, draw] (1) {\scriptsize{1}};
\node at (4.6,2.8) [circle,inner sep=1pt,draw] (2) {\scriptsize{2}};
\node at (6.2,2.6) [circle,inner sep=1pt,draw] (3) {\scriptsize{3}};
\node at (8.7,2.8) [circle,inner sep=1pt,draw] (4) {\scriptsize{4}};
\node at (11.9,2) [circle,inner sep=1pt,draw] (5) {\scriptsize{5}};
\node at (11.9,0) [circle,inner sep=1pt,draw] (6) {\scriptsize{6}};
\node at (8.7,.3) [circle,inner sep=1pt,draw] (7) {\scriptsize{7}};
\node at (6.3,0.4) [circle,inner sep=1pt,draw] (8) {\scriptsize{8}};
\node at (4.3,0.5) [circle,inner sep=1pt,draw] (9) {\scriptsize{9}};
\node at (2.3,0.5) [circle,inner sep=2pt,draw] (10) {\tiny{10}};
\node at (.3,0.5) [circle,inner sep=2pt,draw] (11) {\tiny{11}};

\begin{pgfonlayer}{bg}
\node[inner sep=0pt] (cr) at (-2,2)
    {\includegraphics[width=.25in]{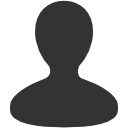}};
\node[inner sep=0pt] (cr) at (-1,2)
    {\includegraphics[width=.35in]{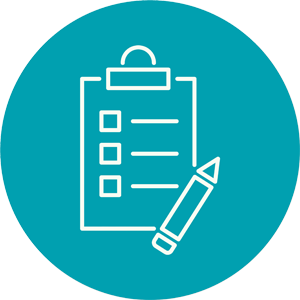}};
\node at (-1,1.2) (b) {\scriptsize Software change};
\node at (-1,0.9) (b) {\scriptsize request};

\node[inner sep=0pt] (init) at (1,2)
    {\includegraphics[width=.35in]{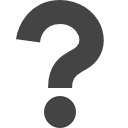}};
\node at (1,2.9) {\scriptsize Initial query};
\node at (1,1.2) (b0) {\scriptsize \bf{(input)}};
%\node at (1,0.7) (b) {\scriptsize (\bf{input})};

\node[inner sep=0pt] (prep) at (3,2)
    {\includegraphics[width=.35in]{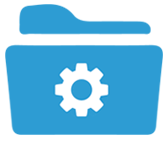}};
\node at (3,1.2) (b) {\scriptsize Preprocessing};
\draw[->,thick] (cr) -- (init);
\draw[->,thick] (init) -- (prep);

\node[inner sep=0pt] (repo) at (4.8,2)
    {\includegraphics[width=.35in]{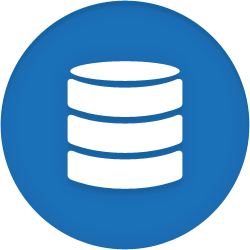}};
%\draw[->,thick] (cr) -- (init);
\node at (5,1.2) (b) {\scriptsize Code search};

\node[inner sep=0pt] (sengine) at (5.4,2)
    {\includegraphics[width=.25in]{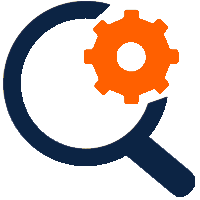}};
\draw[->,thick] (prep) -- (repo);
%\draw[->,thick] (repo) -- (sengine);

\node[inner sep=0pt] (rfb) at (7,2)
    {\includegraphics[width=.30in]{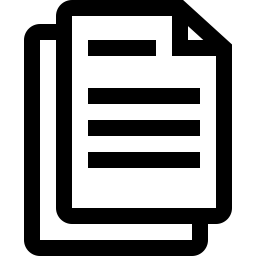}};
\node at (7,3) (b) {\scriptsize Pseudo-relevance};
\node at (7,2.7) (b) {\scriptsize feedback};
\draw[->,thick] (sengine) -- (rfb);

\node[inner sep=0pt] (mine) at (9,2)
    {\includegraphics[width=.35in]{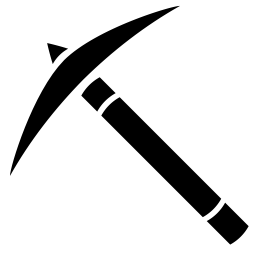}};
\node at (9,1.2) (b) {\scriptsize Source token};
\node at (9,0.9) (b) {\scriptsize selection \& preprocessing};
\draw[->,thick] (rfb) -- (mine);

\node[inner sep=0pt] (crank) at (11,2)
    {\includegraphics[width=.35in]{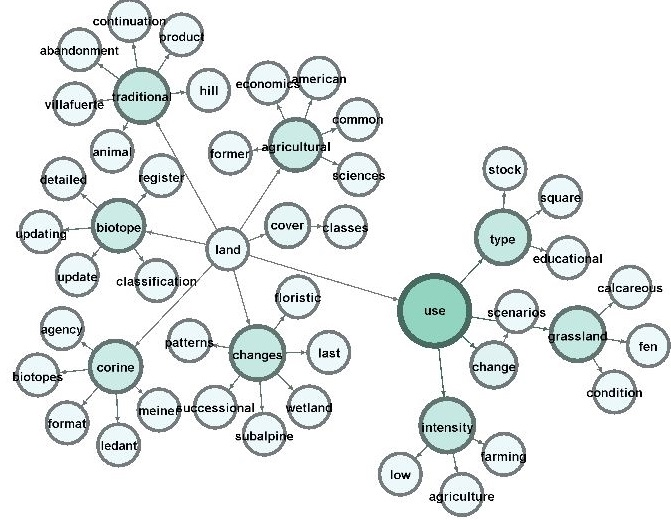}};
\node at (11,3) (b) {\scriptsize Source term graphs for};
\node at (11,2.7) (b) {\scriptsize method and field signatures};
\node[inner sep=0pt] (prank) at (11.7,2)
{\includegraphics[width=.35in]{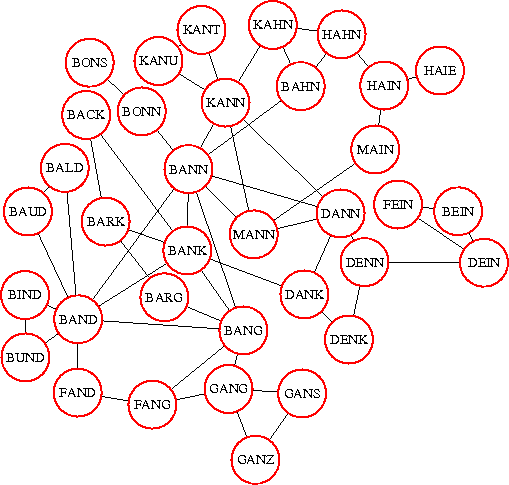}};
\draw[->,thick] (mine) -- (crank);

\node[inner sep=0pt] (calc) at (11,0)
    {\includegraphics[width=.35in]{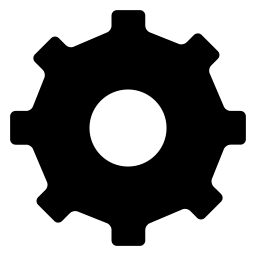}};
\node at (11,-0.8) (b) {\scriptsize CodeRank};
\node at (11,-1.1) (b) {\scriptsize calculation};
\draw[->,thick] (crank) -- (calc);

\node[inner sep=0pt] (ranked) at (9,0)
    {\includegraphics[width=.35in]{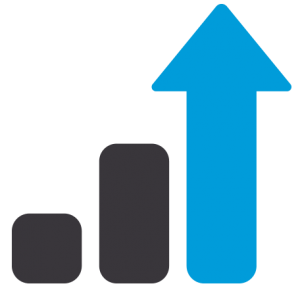}};
\node at (9,-0.8) (b) {\scriptsize Search term};
\node at (9,-1.1) (b) {\scriptsize ranking};
\draw[->,thick] (calc) -- (ranked);

\node[inner sep=0pt] (list) at (7,0)
    {\includegraphics[width=.35in]{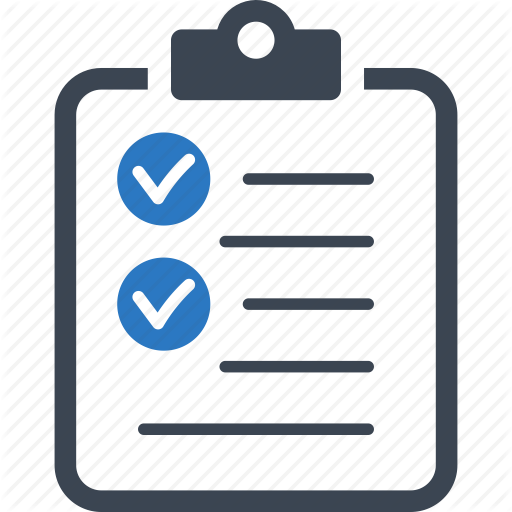}};
\node at (7,-0.8) (b) {\scriptsize Candidate};
\node at (7,-1.1) (b) {\scriptsize reformulations};
\draw[->,thick] (ranked) -- (list);

\node[inner sep=0pt] (sample) at (5,0)
    {\includegraphics[width=.35in]{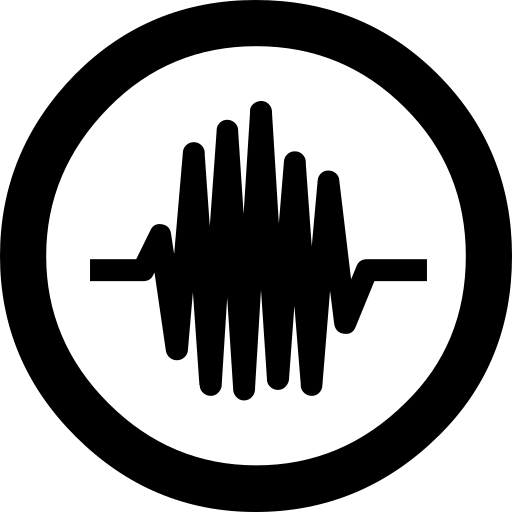}};
%\node[inner sep=0pt] (sett) at (5.3,.5)
%    {\includegraphics[width=.3in]{./sysdiag/setting3.png}};
\node at (5,-0.8) (b) {\scriptsize Quality metric};
\node at (5,-1.1) (b) {\scriptsize data resampling};
\draw[->,thick] (list) -- (sample);

\node[inner sep=0pt] (bestq) at (3,0)
    {\includegraphics[width=.35in]{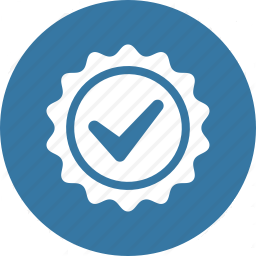}};
\node[inner sep=0pt] (sett) at (3.3,.5)
    {\includegraphics[width=.3in]{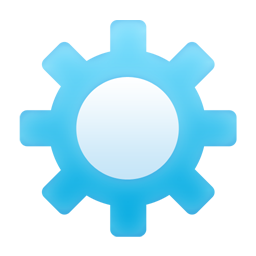}};
\node at (3,-0.8) (b) {\scriptsize Select best};
\node at (3,-1.1) (b) {\scriptsize reformulation};
\draw[->,thick] (sample) -- (bestq);

\node[inner sep=0pt] (merger) at (1,0)
    {\includegraphics[width=.35in]{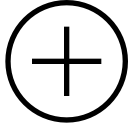}};
\node at (1,-0.8) (b) {\scriptsize Query};
\node at (1,-1.1) (b) {\scriptsize expansion};
\draw[->,thick] (bestq) -- (merger);
\draw[->,thin] (b0) -- (merger);

\node[inner sep=0pt] (expanded) at (-1,0)
    {\includegraphics[width=.35in]{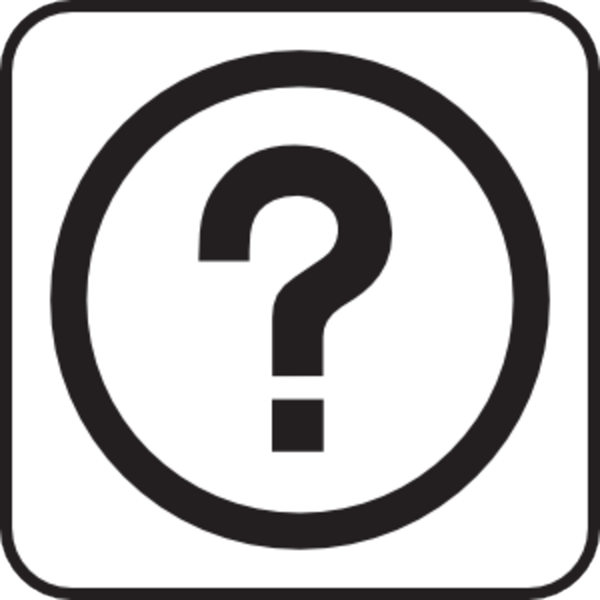}};
\draw[->,thick] (merger) -- (expanded);
\node at (-1,-0.8) (b) {\scriptsize Reformulated query};
\node at (-1,-1.1) (b) {\scriptsize (\textbf{output})};

\end{pgfonlayer}
\end{tikzpicture}
\vspace{-.3cm}
\caption{Schematic diagram of the proposed query reformulation technique--ACER}
\label{fig:sysdiag}
\vspace{-.5cm}
\end{figure*}

\section{ACER: \textbf{A}utomatic Query \textbf{R}eformulation using \textbf{C}odeRank and Docum\textbf{e}nt Structures}\label{sec:acer}

Fig. \ref{fig:sysdiag} shows the schematic diagram of our proposed technique--ACER--for automatic query reformulation.
%Given that software developers often face difficulties in choosing appropriate search terms
%and locating relevant source code for a change request, we reformulate their initial queries by suggesting complementary or relevant search terms from the project source. 
%While traditional approaches mostly rely on TF-IDF based metrics, 
We use a novel graph-based metric of term importance--\emph{CodeRank}-- for source code, and apply source document structures, query quality analysis and machine learning for query reformulation for concept location.
We define CodeRank and discuss different steps of ACER in the following sections. 

\subsection{Pseudo-relevance Feedback}\label{sec:datacoll}
%Although a few studies attempt to collect explicit feedback from the developers,    
%\textbf{Pseudo-relevance Feedback:}
In order to suggest meaningful reformulations to an initial query, feedback on the query is required. 
\citet{gayg} first reformulate queries based on explicit feedback from the developers.  
Although such feedback could be useful, gathering them is often time-consuming and sometimes infeasible.
Hence, a number of recent studies \cite{qsurvey,refoqus,ase2016masud,saner2017masud} apply pseudo-relevance feedback as a feasible alternative.
The top ranked results returned by the code search tool for an initial query are considered as the \emph{pseudo-relevance feedback} for the query. 
We first refine an initial query by removing the punctuation marks, numbers, special symbols and stop words (Step 1, Fig. \ref{fig:sysdiag}).
Then we collect the Top-K (\ie\ $K=10$, best performing heuristic according to our experiments) search results returned by the query, and use them as the source for our candidate terms for query reformulation (Steps 2, 3, Fig. \ref{fig:sysdiag}).

\subsection{Source Token Selection for Query Reformulation}\label{sec:candidate}
%\textbf{Term Proximity and Co-occurrence:} 
%One way of choosing candidate terms for reformulation is to define \emph{context} of query terms within the source code.
%\citet{sisman} apply \emph{spatial code proximity (SCP)} where they choose such terms for query reformulation that frequently co-occur with query terms within a fixed size of window
% in the source code. 
%\citet{ehill} consider the presence of query terms in the method or field signatures 
%as an indicator of thier relevance, and suggest natural language phrases from them as queries for code search.
%That is, both studies return such terms for query reformulation that occur within the \emph{contexts} of the initially chosen query terms.  
%These returned terms could be relevant to the query terms due to their co-occurrences. However, they might not be always the important terms from the document, which might hurt the retrieval performance of the reformulated queries.
%Besides, the quality of the initial query might directly affect the reformulated queries due to the co-occurrence constraints imposed. 
%For example, if the query terms are not chosen carefully and they could not be found in the method or field signatures, \citet{hillicse09} might not perform well.   
%If too generic terms are selected as initial query, \citet{sisman} might also not be able to return specific terms (\ie\ less frequent but important) as reformulation.
%Thus, neither co-occurred nor spatially proximate terms as query context might always help in the effective query reformulation.

\textbf{Global Query Contexts:} Pseudo-relevance feedback on an initial query provides a list of relevant source documents where one or more terms from the query 
generally occur. 
\citet{sisman} choose such terms for query reformulation that frequently co-occur with the initial query terms within a fixed window size
 in the feedback documents.
\citet{hillicse09} consider presence of the query terms in method signatures or field signatures 
as an indicator of their relevance, and suggest natural language phrases from them as reformulated queries.
Both reformulation approaches are highly subject to the quality of the initial query due to their imposed constraints-- co-occurrences with query terms \cite{sisman} and textual similarity with query terms \cite{hillicse09}.   
\citet{rocchio} determines importance (\ie\ TF-IDF) of a candidate term across all the feedback documents, and suggests the top-ranked terms for query reformulation.
\citet{carmel} suggest that a single natural language query might focus on multiple topics, and 
different parts of the returned results might cover different topics.
That is, the same candidate term is not supposed to be important across all the feedback documents. 
In other words, accumulating term weight across all the documents might not always return the most appropriate terms for query reformulation.
Such sort of calculation might add unnecessary noise to the term weight from the unrelated topics.  
Hence, we consider all the feedback documents as a \emph{single body of structured texts} which acts as a \emph{``global context''} for the query terms. Thus, with the help of an appropriate term weighting method, the terms representing the most dominant topic across the feedback documents (\ie\ also in the initial query) could simply stand out, and could be chosen for reformulation. 
%In other words, we let the relevance feedback \emph{``speak for itself''}, and choose the dominant terms from the global context of the query for reformulation. 

\textbf{Candidate Token Mining:}
%Furthermore, we preserve the level of developers' intent behind the source code, and extract two different query contexts by collecting method and field signatures from all the documents provided as pseudo-relevance feedback.
Developers often express their intent behind the code and encode domain related concepts
in the identifier names and comments \cite{domainterm}. 
However, code comments are often inadequate or outdated \cite{codes}. All identifier types also do not have the same level of importance. For example, while the signature of a method encodes the high level intent for the method, its body focuses on granular level implementation details and thus possibly contains more noisy terms \cite{hillicse09}. 
In fact, \citet{hillicse09} first analyze method signatures and field signatures to suggest natural language phrases as queries for code search.    
%While most existing studies consider all the terms of comments and identifier names from the code,
%we focus on gathering only semantically rich words for our task.    
In the same vein, we thus also consider method signatures ($msig$) and field signatures ($fsig$) as the source for our candidate reformulation terms.
% whileour query context being much wider rather than a single document.
%It should not noted that we do not impose co-occurrence constraints (with the query terms) on the candidate terms.
%our method of term weighting (\ie\ determining term importance) being totally novel. 
We extract structured identifier names from these signatures using appropriate regular expressions \cite{rigby} (Step 4, Fig. \ref{fig:sysdiag}).
Since different contexts of a source document might convey different types or levels of semantics (\ie\ developers' intent), 
we develop a separate candidate token set ($CT_{sig}$) for each of the two signature types ($sig\in\{msig,fsig\}$) from the feedback documents ($\forall d\in D_{RF}$) as follows:
%The phrasal concept of \citeauthor{hillicse09} is based on such co-occurrences.
%Thus, they could be considered as an equivalent to natural language phrases in the source code \cite{ehill,hillicse09}. 
%Such phrases often capture and encode certain concepts from the problem domain \cite{tidier} which could be leveraged for determining importance of the corresponding terms. 
%We thus apply a set of regular expressions, and extract all the structured tokens from the source code (Step 4, Fig. \ref{fig:sysdiag}).  
\begin{equation*}\label{eq:candidate}
\setlength\abovedisplayskip{2pt}
\setlength\belowdisplayskip{2pt}
CT_{sig}=\bigcup\limits_{\forall d\in D_{RF}}\{\exists t \in T_{sig} \}\mid structured(t)\wedge T_{sig}= sig(d)
\end{equation*}
Here $sig(d)$ extracts all tokens from method signatures or field signatures, and $structured(t)$ determines whether the token $t\in T_{sig}$ is structured or not.
Although we deal with Java source code in this research where the developers generally use camel case tokens (\eg\ \texttt{MessageType})  or occasionally might use same case tokens (\eg\ \texttt{DECIMALTYPE}), our approach can be easily replicated for snake case tokens (\eg\ \texttt{reverse_traversal}) as well.

\subsection{Source Code Preprocessing}\label{sec:preprocess}
\textbf{Token Splitting:} 
Structured tokens often
%may take one of these forms-- camel case , snake case or simple concatenation . They 
consist of multiple terms where the terms co-occur (\ie\ are concatenated) due to their semantic or temporal relationships \cite{shepherd}.
%Once structured tokens are extracted from the source code, 
%We perform natural language preprocessing on them 
%The underlying idea is not only to extract the salient terms (\ie\ related to problem domain) from the source code but also to decode and capture the inferred relationships among them using their contexts. 
We first split each of the complex tokens based on punctuation marks (\eg\ dot, braces) which returns the individual tokens (Step 4, Fig. \ref{fig:sysdiag}).
Then each of these tokens is splitted using a state-of-the-art token splitting tool--\emph{Samurai} \cite{samurai}--given that regular expression based splitting might not be always sufficient enough. \emph{Samurai}
% is reported to perform well not only with camel case tokens but also with the same case complex tokens. It 
mines software repositories to identify the most frequent terms, and then suggests the splits for a given token.
We implement \emph{Samurai} in our working environment where our subject systems (Section \ref{sec:dataset}) are used for mining the frequent terms, and the author's provided prefix and suffix lists \cite{prefix} are applied to the splitting task.  
%Thus, the tokens that neither have any structure of interest nor can be splitted using the state-of-the-art tool are kept away from further analysis.        

\textbf{Stop word and Keyword Removal:}
%Stop words are the words that frequently occur in the regular texts but convey very little or no semantics \cite{wordsim}. 
Since our structured tokens comprise of natural language terms, we discard stop words from them as a common practice (Step 4, Fig. \ref{fig:sysdiag}).
We use a standard list \cite{stopword} hosted by Google for stop word removal.    
Programming keywords can often be considered as the equivalence of stop words in the source code which are also discarded from our analysis.
Since we deal with Java source code, the keywords of Java are considered for this step.  
As suggested by earlier study \cite{refoqus}, we also discard insignificant source terms (\ie\ having word length$<3$) from our analysis.

\textbf{Stemming:} It extracts the root (\eg\ ``send") out of a word (\eg\ ``sending"). 
Although
% stemming is widely applied to regular texts for better performance in the information retrieval, 
existing studies suggest contradictory \cite{kevic,saha} or conflicting \cite{stemming} evidences for stemming with 
the source code,
%Besides, stemming could be sensitive to the search query length \cite{stemming}.    
%Besides, our preliminary investigation also reported negative findings on the stemming.   
%We thus generally avoid stemming on source terms in our experiments.
%However, 
we  investigate the impact of stemming with RQ$_4$ where Snowball stemmer \cite{stemming,prompter} is used for stemming.

\subsection{Source Term Graph Development}
Once candidate tokens are extracted from method signatures and field signatures, and are splitted into candidate terms, we develop source term graphs (\eg\ Fig. \ref{fig:tgraph}) from them (Step 5, Fig. \ref{fig:sysdiag}).  
Developers often encode their intent behind the code and domain vocabulary into the carefully crafted identifier names where multiple terms are concatenated.
%Structured tokens encode majority of the semantics of the code, and thus could be exploited for search query reformulation. 
For example, the method name--\texttt{getChatRoomBots}--looks like a natural language phrase--\emph{``get chat room bots''}--when splitted properly.
Please note that each of these three terms--\emph{``chat'', ``room'' } and \emph{``bots''}-- co-occur with each other to convey an important concept-- a robotic technology, 
and thus, they are semantically connected. On the other hand, the remaining term--\emph{``get''}-- co-occurs with them due to a temporal relationship (\ie\ develops a verbal phrase).    
Similar phrasal representations (refined with lexical matching) were directly returned by \citeauthor{hillicse09} for query reformulation.
%where the developers were expected to
%However, as argued previously, collecting explicit feedback from the developers could be time-consuming and infeasible.7
However, their approach could be limited due to the added constraint (\eg\ warrants query terms in signatures).
%However, given the limitations due to added constraint (\eg\ warrants query terms in the signatures), 
We thus perform further analysis on such phrases, and exploit the co-occurrences among the terms for our graph based term weighting. 
In particular, we encode the term co-occurrences into connecting edges ($E$) in the term graph ($G(V,E)$) 
where the individual terms ($V_i$) are denoted as vertices ($V$).  
\begin{equation*}\label{eq:candidate}
\setlength\abovedisplayskip{2pt}
\setlength\belowdisplayskip{2pt}
V=\bigcup\limits_{\forall t\in CT_{sig}}\{V_{i}\in splitted(t)\mid validterm(V_i)\}
\end{equation*}
\begin{equation*}
\setlength\abovedisplayskip{2pt}
\setlength\belowdisplayskip{2pt}
E=\bigcup\limits_{\exists V_i,V_j\in V}\{(V_i,V_j)\mid V_i,V_j\in t~ \wedge\mid i-j\mid =1\}  
\end{equation*}
Here $splitted(t)$ returns individual terms from the token $t\in CT_{sig}$, and $validterm(V_i)$ determines whether the term is valid (\ie\ not an insignificant or a stop word) or not.
We consider a \emph{window size} of \emph{two} within each phrase for capturing co-occurrences among the terms. Such window size for co-occurrence was reported to perform well by the earlier studies \cite{rada,blanco,saner2017masud}.
Thus, the above method name can be represented as the following edges-- \emph{get}$\longleftrightarrow$\emph{chat}, \emph{chat}$\longleftrightarrow$\emph{room}, and \emph{room}$\longleftrightarrow$\emph{bots} -- in the term graph.
%\emph{happen}$\longleftrightarrow$\emph{version}, and \emph{version}$\longleftrightarrow$\emph{installed}. 
%We then encode such relationships into the connecting edges among the corresponding nodes in the graph. 
That is, if a set of terms are frequently shared across multiple tokens from two signature types, such occurrences are represented as the high connectivity in the term graph (\eg\ \emph{``Classpath"} in Fig. \ref{fig:tgraph}).

\subsection{CodeRank Calculation}
\label{sec:textrank}

%\citet{pagerank} first propose \emph{PageRank} algorithm for web link analysis which identifies reputed web pages based on the recommendations (\ie\ incoming links) from other reputed pages. 
%%That is, a web page is considered important by PageRank only if it is referred (\ie\ recommended) by other important pages over the web.
%%Google uses PageRank for page indexing and web search.
%%Such scoring method is highly effective and reliable for web search, as was demonstrated by Google's success. 
%\citet{rada} first adapt PageRank \cite{pagerank} for regular text documents as \emph{TextRank} where they represent 
%individual terms as vertices and
%co-occurrences among the terms (within a sliding window) as the connecting edges in the text graph.
%\citet{saner2017masud} apply TextRank to bug reports to suggest the initial query for concept location.

\textbf{CodeRank:} 
%Once candidate terms and their relationships are represented as a source term graph, we determine their relative importance using the topological characteristics of the graph.  
PageRank \cite{pagerank} is one of the most popular algorithms for web link analysis which was later adapted by \citet{rada} for text documents as TextRank.
%\citet{saner2017masud} apply TextRank to bug reports for suggesting initial queries for concept location.
In this research, we adapt our term weighting method from TextRank \cite{rada,blanco,saner2017masud} for source code, and we call it \emph{CodeRank}.
%Since we estimate relative weight for terms from the source code, 
%While the underlying scoring technique is same as TextRank, CodeRank is a novel term weighting method for source code. 
To date, only traditional term weights (\eg\ TF, TF-IDF \cite{rocchio,refoqus,sisman}) are applied to source code which were originally proposed for regular texts \cite{tfidf} and are mostly based on isolated frequencies.
On the contrary, CodeRank not only analyzes the connectivity (\ie\ incoming links and outgoing links) of each source term, but also the relative weight  
of the connected terms
from  the graph recursively, and calculates the term weight, $S(V_{i})$, as follows (Step 6, Fig. \ref{fig:sysdiag}):
\begin{equation*}\label{eq:textrank}
\setlength\abovedisplayskip{2pt}
\setlength\belowdisplayskip{2pt}
S(V_{i})=(1-\psi)+\psi\sum_{j\epsilon In(V_{i})}\frac{S(V_{j})}{|Out(V_{j})|}~~ (0 \le \psi \le1)
\end{equation*}
\noindent
Here, $In(V_{i})$, $Out(V_{j})$, and $\psi$ denote the vertices to which $V_i$ is connected through incoming links, the vertices to which $V_j$ is connected through outgoing links, and the damping factor respectively. As shown earlier using the example--\texttt{getChatRoomBots}, co-occurred terms complement each other with their semantics which are represented as bi-directional edges in the term graph. Thus, each ($V_i$) of the vertices from the graph has equal number of incoming links and outgoing links, \ie\ \emph{in-degree($V_i$)=out-degree($V_i$)}.    
%In the \emph{text graph}, each of the edges is bi-directional (\ie\ terms depend on each other), and thus \emph{in-degree} is equal to the \emph{out-degree} for a node (\ie\ term).

\textbf{Parameters and Configurations:}
\citet{pagerank} consider damping factor, $\psi$, as the probability of randomly choosing a web page in the context of web surfing by a random surfer.
That is, $1-\psi$ is the probability of jumping off that page by the surfer.
They use a well-tested value of 0.85 for $\psi$ which was later adopted by \citet{rada} for text documents.
Similarly, we also use the same value of $\psi$ for \emph{CodeRank} calculation. 
Each of the vertices is assigned to a default value (\ie\ base term weight) of 0.25 (as suggested by earlier studies \cite{pagerank,rada}) with which CodeRank is calculated.
It should be noted that the base weight of a vertex does not determine its final weight when PageRank based algorithms are applied \cite{rada}. CodeRank adopts the underlying mechanism of recommendation or votes \cite{saner2017masud,rada} for term weighting. That is, each vertex feeds off from the scores of surrounding connected vertices from the graph in terms of recommendation (\ie\ incoming edges).     
%We initialize each of the terms in the graph with a default value of 0.25,
%and run an iterative version of the algorithm \cite{pagerank}. It should be noted that the initial value of a term does not affect its final score \cite{rada}.
PageRank generally has two modes of computation--\emph{iterative} version and \emph{random walk} version.
We use the iterative version for CodeRank, and the computation iterates until the weights of the terms converge below a certain threshold or they reach the maximum iteration limit (\ie\ 100 as suggested by \citet{blanco}). 
As applied earlier \cite{rada}, we apply a heuristic threshold of 0.0001 for the convergence checking. The algorithm captures importance of a source term not only by estimating its local impact but also by considering its global influence over other terms.  
For example, the term, \emph{``Classpath''}, Fig. \ref{fig:tgraph}, occurs in multiple structured tokens (Listing \ref{lst:example}), complements the semantics of five other terms, and thus is highly important within the term graph (\ie\ Fig. \ref{fig:tgraph}).
%\emph{TextRank} applies the underlying mechanism of a recommendation (\ie\ voting) system, where a term recommends (\ie\ votes) another term if the second term complements the semantics of the first term in any way \cite{rada}. 
%The algorithm captures recommendation for a term in terms of incoming links in the \emph{text graph} (\eg\ Fig. \ref{fig:tgraph}) from another terms both in local (\ie\ same sentence) and global (\ie\ entire document) context, and determines importance of the term.
Once the iterative computation is over, each of the terms from the graph is found with a numeric score. We consider these scores as  the relative weight or importance of the corresponding terms from the source code.

\subsection{Suggestion of the Best Query Reformulation}\label{sec:bestref}
\textbf{Candidate Reformulation Selection:}
Algorithms \ref{acer} and \ref{gc} show the pseudo-code of our query reformulation technique--ACER--for concept location. 
We first collect pseudo-relevance feedback for the initially provided query ($Q$) where Top-K source documents are returned (Lines 3--5, Algorithm \ref{acer}).
Then we collect method signatures and field signatures from each of the documents ($\forall d\in D_{RF}$), 
and extract structured tokens from them.
We prepare three token sets--$CT_{msig}, CT_{fsig}$ and $CT_{comb}$ from these signatures (Lines 6--12, Algorithm \ref{acer}, Step 4, Fig. \ref{fig:sysdiag}) where $CT_{comb}$ combines tokens from both signatures.  
Then we perform limited natural language preprocessing on each token set where \emph{Samurai} algorithm \cite{samurai} is used for token splitting.  
We develop separate term graph for each of these token sets where individual terms are represented as vertices, and term co-occurrences are encoded as connecting edges (Lines 3--7, Algorithm \ref{gc}, Step 5, Fig. \ref{fig:sysdiag}).    
We apply CodeRank term weighting to each of the graphs which provides a ranked list of terms based on their relative importance.
Then we select Top-K (\eg\ $K=10$) important terms from each of the three graphs, and prepare three reformulation candidates (Lines 8--12, Algorithm \ref{gc}, Steps 6, 7, 8, Fig. \ref{fig:sysdiag}).

\setlength{\intextsep}{2pt}
\begin{algorithm}[!h]
\caption{ACER: Proposed Query Reformulation}
\label{acer}
\small
\begin{algorithmic}[1]
\Procedure{ACER}{$Q$}\Comment{$Q$: initial search query}
\State $L \gets$ \{\}\Comment{list of best reformulation query terms}
\LineComment{collecting pseudo-relevance feedback for $Q$}
\State $Q_{pp}\gets$ preprocess($Q$)
\State $D_{RF} \gets$ getRelevanceFeedback($Q_{pp}$)
\LineComment{collecting candidate source tokens from signatures}
\For{SourceDocument $d\in D_{RF}$}
\State $CT_{msig}\gets CT_{msig}~\cup$ getMethodSigTokens($d$) 
\State $CT_{fsig}\gets CT_{fsig}~\cup$ getFieldSigTokens($d$) 
\EndFor
\State $CT_{comb}\gets CT_{msig}\cup CT_{fsig}$ 
\State $CT_{all}\gets \{CT_{msig}, CT_{fsig}, CT_{comb}\}$
\For{TokenList $CT_{sig}\in CT_{all}$}
\State $QR[sig]\gets$ getQRCandidate($CT_{sig}$)
\EndFor
%\LineComment{selecting the best reformulation candidate}
\LineComment{suggesting the best reformulated query for $Q$}
\State $QD\gets$ resample(getQueryQualityMetrics($QR$))
\State $QR_{best}\gets$ getBestCandidateUsingML($QR, Q_{pp}, QD$)
%\LineComment{suggesting the best reformulated query for $Q$}
\State $L \gets$ combine($Q_{pp},QR_{best}$)
\State \textbf{return} $L$
\EndProcedure
\end{algorithmic}
\end{algorithm}
\setlength{\floatsep}{0pt}

\textbf{Selection of the Best Reformulation:}
%Query reformulations generally involve addition, deletion or replacement of certain terms from a given query.  
\citet{refoqus} argue that the same type of reformulation (\ie\ addition, deletion or replacement of query terms) might not be appropriate for all given queries. In the same vein, we argue that 
query reformulations from different contexts of the source document (\eg\ method signature, field signature) might have different level of effectiveness given that they embody different level of semantics and noise.
%From our experiments, we also note that the candidate 
That means, one or more of the reformulation candidates could improve the initial query, but the best one should be chosen carefully for useful recommendation.   
%Once CodeRank term weight is calculated for each of the candidate reformulation terms, we rank them based on their relative importance.
%Then we choose Top-K important terms as a candidate reformulation ($qr\in QR$) from each of the graphs (Lines 22--26, Algorithm \ref{acer}, Steps 6, 7, Fig. \ref{fig:sysdiag}).
%It should be noted that we select three reformulation candidates for each query--$QR_{msig}, QR_{fsig}$, and $QR_{comb}$. 

%\vspace{-.3cm}
\citet{qperf} suggest that quality of a query with respect to the corpus could be determined using four of its statistical properties-- \emph{specificity, coherency, similarity} and \emph{term relatedness}--that comprise of 21 metrics \cite{carmelbook}.
They apply machine learning on these properties, and separate high quality queries from low quality ones.   
%Given that we collect multiple reformulation candidates from various contexts of the source document
We thus also similarly apply machine learning on our reformulation candidates (and their metrics), and develop classifier model(s) where \emph{Classification And Regression Tree} (CART) is used as the learning algorithm \cite{qperf}. 
Since only the best of the four reformulation candidates (\ie\ including baseline) is of our interest, the training data was inherently skewed.  
We thus perform \emph{bootstrapping} (\ie\ random resampling) \cite{resample,bootstrapping} on the data multiple times (\eg\ 50) with 100\% sample size and replacement (Step 9, Fig. \ref{fig:sysdiag}), train multiple models using the sampled data, and then record their output predictions.
Then, we average all the predictions for each test instance from all models, and determine their average probability of being the best candidate reformulation. Thus, we identify the best of the four candidates using our models, and suggest the best reformulation to the initial query (Lines 16--20,  Algorithm \ref{acer}, Steps 10, 11, Fig. \ref{fig:sysdiag}). 
\citet{twkraft} suggest that repetition of certain query terms might improve retrieval performance of the query.
If none of the candidates is likely to improve the initial query according to the quality model (\ie\ baseline itself is the best), we repeat all the terms from the initial query as the reformulation.   

\setlength{\intextsep}{2pt}
\begin{algorithm}[!h]
\caption{getQRCandidate: Get a candidate reformulation}
\label{gc}
\small
\begin{algorithmic}[1]
\Procedure{getQRCandidate}{$CT_{sig}$}\Comment{$CT_{sig}$: extracted candidate tokens from the signatures $sig$}
\State $QR_{sig}\gets$ \{\} \Comment{candidate query reformulation}
\LineComment{extracting terms and their co-occurrences}
\State $ST_{sig}\gets$ preprocess(Samurai($CT_{sig}$))
\State $CO_{sig}\gets$ getTermCo-occurrences($ST_{sig}$, $CT_{sig}$)  
\LineComment{developing term graph from token set}
\State $G_{sig} \gets$ developTermGraph($ST_{sig},CO_{sig}$)
%\EndFor
\LineComment{calculating CodeRank using the graph}
\State $CR_{sig} \gets$ normalize(calculateCodeRank($G_{sig}$))
\LineComment{getting candidate reformulated query}
\State $QR_{sig}\gets$ getTopKTerms(sortByValue($CR_{sig}$))
\State \textbf{return} $QR_{sig}$
\EndProcedure
\end{algorithmic}
\end{algorithm}
\setlength{\intextsep}{0pt}

%We thus append the preprocessed initial query ($Q_{pp}$) in the reformulated query as well.
%\citet{carmel} suggest that \emph{Jensen Shannon Divergence (JSD)} between the retrived documents of a query and the corpus documents could be used as a proxy to the \emph{average precision} of the query. That is, the more separable the relevant documents are from rest of the corpus, the better is the quality of the query.

\textbf{Working Example:} Let us consider the query--\{debugger source lookup work variables\}--from our running example in Table \ref{table:we}. 
Our term weighting method--\emph{CodeRank}--extracts three candidate reformulations from method signatures and field signatures. 
We see that different candidates have different level of effectiveness (\ie\ rank 02 to rank 16), and in this case, the candidate from the method signatures ($QR_{msig}$) is the most effective.
Our technique--ACER-- not only prepares such candidate queries from various contexts (using a novel term weighting method) but also suggests the best candidate ($QR_{best}$) for query reformulation. 
The reformulated query--\{debugger source lookup work variables \emph{launch debug resolve required classpath}\}  -- returns the first correct result at the top position (\ie\ rank 02) of the result  list which is highly promising.
Such effective reformulations are likely to reduce a developer's effort during software change implementation.

\begin{table}[!t]
\centering
\caption{A Working Example (Bug \#31110, eclipse.jdt.debug)}\label{table:we}
\vspace{-.2cm}
\resizebox{3.5in}{!}{%
\begin{threeparttable}
\begin{tabular}{l| p{6.5cm}|c}
\hline
\textbf{Source} & \textbf{Query Terms} & \textbf{QE}\\
\hline
\hline
Bug Title & Debbugger Source Lookup does not work with variables & 72 \\
\hline
Initial   &\multirow{2}{*}{\{debugger source lookup work variables\}} & 77\\
\hhline{~~~}
Query ($Q$) & &  \\
\hline
$Q'_{msig}$ & $Q_{pp}~\cup~$($QR_{msig}$=\{launch debug resolve required classpath\})  &  \textbf{02}\\
\hline
$Q'_{fsig}$ & $Q_{pp}~\cup~$($QR_{fsig}$=\{label classpath  system  resolution launch\}) & \textbf{06} \\
\hline
$Q'_{comb}$ & $Q_{pp}\cup~$($QR_{comb}$=\{java type launch classpath label\})& 16 \\
\hline
\hline
\multicolumn{3}{c}{$QR_{best} = $ getBestCandidateUsingML($QR_{msig}$, $QR_{fsig}$, $QR_{comb}$, $Q_{pp}$, $QD$) }  \\
\hline
\hline
$Q'_{ACER}$ & $Q_{pp}\cup ~QR_{best}$    &  \textbf{02} \\
\hline
\end{tabular}
%\begin{tablenotes}
%\item [1] No. of example pairs for which relative quality evaluation matches with that of StackOverflow
%\item [2] \% of agreement, \item [3] \% of disagreement
\centering
\textbf{QE} = Query Effectiveness, rank of the first correct result returned by the query
 %\end{tablenotes}
\end{threeparttable}
}
%\vspace{-.2cm}
\end{table}

%\begin{table*}
%\centering
%\caption{Experimental Results}\label{table:result}
%\resizebox{5.5in}{!}{%
%\begin{threeparttable}
%\begin{tabular}{l|c|c|c||c|c|c||c}
%\hline
%\multirow{2}{*}{\textbf{Metric}} & \multicolumn{3}{c||}{\texttt{Log4j}} & \multicolumn{3}{c||}{\texttt{eclipse.jdt.debug}} & \textbf{Average}\\
%\hhline{~-------}
%& $\mathbf{T_3}\tnote{1}$ & $\mathbf{T_4}\tnote{2}$ & $\mathbf{T_5}$\tnote{3} & $\mathbf{T_3}$ & $\mathbf{T_4}$ & $\mathbf{T_5}$ & $\mathbf{T_5}$  \\
%\hline
%No. of Tasks Solved (NTS) & 86(230) & 98(230) & \textbf{111}(230) & 48(119) & 54(119) & \textbf{60}(119) & -- \\
%\hline
%\% of Tasks Solved (PTS) & 37.39\% & 42.61\% & \textbf{48.26}\% & 40.34\% & 45.38\% & \textbf{50.42}\% & 49.34\%\\
%\hline
%Mean Average Precision (MAP) & \textbf{66.54}\% & 62.82\% & 61.16\% & \textbf{56.79}\% & 51.45\% & 53.16\% & 57.16\% \\
%\hline
%Mean Recall (MR) & 53.22\% & \textbf{55.54}\% & 54.66\% & \textbf{73.56}\% & 73.27\% & 72.00\% & 63.33\% \\
%\hline
%\end{tabular}
%%\begin{tablenotes}
%%\item [1] No. of example pairs for which relative quality evaluation matches with that of StackOverflow
%%\item [2] \% of agreement, \item [3] \% of disagreement
%\centering
% $^1$Results for three search terms,  $^2$Results for four search terms,  $^3$Results for five search terms
% %\end{tablenotes}
%\end{threeparttable}
%%\vspace{-.2cm}
%}
%\vspace{-.5cm}
%\end{table*}

\section{Experiment}
\label{sec:experiment}
Although pre-retrieval methods (\eg\ coherency, specificity \cite{qperf}) are lightweight and reported to be effective for query quality analysis, post-retrieval methods are more accurate and more reliable \cite{refoqus}. 
Existing studies \cite{refoqus,saner2017masud,saha,trconfig} also adopt these methods widely for evaluation and validation. 
We evaluate our term weighting method and query reformulation technique using 1,675 baseline queries and three performance metrics. We also compare our technique with five closely related existing techniques \cite{rocchio,qsurvey,sisman,hillicse09,refoqus}.
We thus answer five research questions using our experiments as follows:
\begin{itemize}
\item \textbf{RQ$\mathbf{_1}$:} Does query reformulation of ACER improve the baseline queries significantly in terms of query effectiveness and retrieval performance?
\item \textbf{RQ$\mathbf{_2}$:} Does CodeRank perform better than traditional term weighting methods (\eg\ TF, TF-IDF) in identifying effective search terms from the source code?
\item \textbf{RQ$\mathbf{_3}$:} Does employment of document structure improve ACER's suggestion on good quality search terms from the source code?   
\item \textbf{RQ$\mathbf{_4}$:} How stemming, query length, and relevance feedback size affect the performance of our technique?       
\item \textbf{RQ$\mathbf{_5}$:} Can ACER outperform the existing query reformulation techniques from the literature in terms of effectiveness and retrieval performance of the queries?
\end{itemize}

\begin{table}[!t]
\centering
\caption{Experimental Dataset}\label{table:dataset}
\vspace{-.2cm}
\resizebox{3.5in}{!}{%
\begin{threeparttable}
\begin{tabular}{l|c|c||l|c|c}
\hline
\textbf{System} & \textbf{\#Classes}  & \textbf{\#CR} & \textbf{System}& \textbf{\#Classes} & \textbf{\#CR} \\
\hline
\hline
\texttt{eclipse.jdt.core}--4.7.0 & 5,908  & 198 & \texttt{ecf}--279.279 &  2,827 & 154 \\
\hline
\texttt{eclipse.jdt.debug}--4.6.0 & 1,519  & 154 & \texttt{log4j}--1.2.18 & 309  & 28  \\
\hline
\texttt{eclipse.jdt.ui}--4.7.0 & 10,927 & 309 & \texttt{sling}--9.0 & 4,328 & 76 \\
\hline
\texttt{eclipse.pde.ui}--4.6.0 & 5,303 & 302  & \texttt{tomcat70}--7.0.73 & 1,841 & 454 \\
\hline
\end{tabular}
%\begin{tablenotes}
%\item [1] No. of example pairs for which relative quality evaluation matches with that of StackOverflow
%\item [2] \% of agreement, \item [3] \% of disagreement
\centering
\textbf{CR}= Change requests
% %\end{tablenotes}
\end{threeparttable}
}
\vspace{-.4cm}
\end{table}

\begin{table*}[!t]
\centering
\caption{Effectiveness of ACER Query against Baseline Query}\label{table:effectiveness}
\vspace{-.2cm}
\resizebox{7.2in}{!}{%
\begin{threeparttable}
\begin{tabular}{l|c|c|c|c|c|c|c|c||c|c|c|c|c|c|c||c}
\hline
\multirow{2}{*}{\textbf{System}} & \multirow{2}{*}{\textbf{\#Queries}} &\multicolumn{7}{c||}{\textbf{Improvement}} & \multicolumn{7}{c||}{\textbf{Worsening}} & \textbf{Preserving}   \\
%\hhline{~~---------------}
%\textbf{Metric}  & \multicolumn{2}{c|}{\textbf{Class Name}}  & \multicolumn{2}{c|}{\textbf{Method Name}} & \multicolumn{2}{c|}{\textbf{Variable Name}} & \multicolumn{2}{c||}{\textbf{All Combined}}  & \multicolumn{2}{c|}{\textbf{Class Name}}  & \multicolumn{2}{c|}{\textbf{Method Name}} & \multicolumn{2}{c|}{\textbf{Variable Name}} & \multicolumn{2}{c}{\textbf{All Combined}}\\
\hhline{~~---------------}
& & \#Improved & Mean & Q1 & Q2 & Q3 & Min. & Max. & \#Worsened & Mean & Q1 & Q2 & Q3 & Min. & Max. & \#Preserved\\
\hline
\hline
\textbf{ecf} & 154 & 100 \textbf{(64.94\%)} & 71 & \textbf{8} & 20 &  58 & 1 & 654 & 	5 (3.25\%) & 125 & 48 & 88 & 220 & 43 & 329 & 49 (31.82\%) \\
\hline
\textbf{jdt.core}  & 198 & 125 \textbf{(63.13\%)} & 89 & \textbf{8} &  20 & 51 & 1 & 1,485 &	 7 (3.54\%) & 72 & 16 & 38 & 132 & 13 & 195 & 66 (33.33\%)\\
\hline
\textbf{jdt.debug}  & 154 & 110 \textbf{(71.43\%)} & 72 & \textbf{10} &  23 & 73 & 1 & 1,234 &	3 (1.95\%) & 138 & 48 & 102 & 265 & 48 & 265 & 41 (26.62\%)\\
\hline
\textbf{jdt.ui} & 309 & 216 \textbf{(69.90\%)} & 169 & \textbf{10} &  27 & 92 & 1 & 3,162 &	 13 (4.21\%) & 254 & 39 & 91 & 368 & 19 & 1,369 & 80 (25.89\%)\\
\hline
\textbf{pde.ui}  & 302 & 191 \textbf{(63.25\%)} & 143 & \textbf{8} &  33 & 102 & 1 & 2,304 & 	7 (2.32\%) & 507 & 70 & 477 & 1,060 & 40 & 1,172 & 104 (34.44\%)\\
\hline
\textbf{log4j}  & 28 & 23 \textbf{(82.14\%)} & 35 &  \textbf{12} &  17 & 58 & 3 & 136 &	 0 (0.00\%) & - & - & - & - & - & - & 5 (17.86\%)\\
\hline
\textbf{sling}  & 76 & 59 \textbf{(77.63\%)} & 165 & \textbf{9} &  18 & 120 & 2 & 1,940 &	 0 (0.00\%) & - & - & - & - & - & - & 17 (22.37\%)\\
\hline
\textbf{tomcat70}  & 454 & 345 \textbf{(75.99\%)} & 236 & 21 &  92 & 291 & 1 & 1,675 & 22 (4.84\%) & 292 & 97 & 261 & 429 & 34 & 938 & 87 (19.16\%)\\
\hline
 & Total = 1,675 & \textbf{Avg = 71.05\%}  & &  &  &  &  &  & Avg = 2.51\%  &  &  &  &  &  &  & \textbf{Avg = 26.44\%}    \\
\hline
%\textbf{Recall} & 36.91\% & 59.18\% & 38.40\% & 59.25\% & 38.36\% & 59.85\% & 42.70\% & \textbf{62.83}\% & 37.50\% & 55.34\% & 39.85\% & 58.20\% & 39.78\% & 57.91\% & 43.03\% & \textbf{61.54}\%\\
%\hline
\end{tabular}
%\begin{tablenotes}
%\item [1] No. of example pairs for which relative quality evaluation matches with that of StackOverflow
%\item [2] \% of agreement, \item [3] \% of disagreement
 %\end{tablenotes}
\centering
\textbf{jdt.core} = \texttt{eclipse.jdt.core}, \textbf{jdt.debug} = \texttt{eclipse.jdt.debug}, \textbf{jdt.ui} = \texttt{eclipse.jdt.ui}, \textbf{pde.ui} = \texttt{eclipse.pde.ui}, \textbf{Mean} = Mean rank of first correct results returned by the queries, \textbf{Q$_i$}=  $i^{th}$ quartile of all ranks considered
\end{threeparttable}
%\vspace{-.2cm}
}
\vspace{-.7cm}
\end{table*}
 
\subsection{Experimental Dataset}\label{sec:dataset}
\label{sec:dataset}
\textbf{Data Collection:} We collect a total of 1,675 bug reports from eight open source subject systems (\ie\ five \emph{Eclipse} systems and three \emph{Apache} systems) for our experiments.
Table \ref{table:dataset} shows the experimental dataset.
%We first extract all the resolved (\ie\ marked as RESOLVED) bug reports of each subject system from BugZilla and JIRA repositories submitted within a reasonable time interval (Oct, 2001 -- Aug, 2016). 
%Then we consult with the version control history of each system from GitHub, analyze their commit messages, and look for specific Bug ID in those messages \cite{bugid}.
%We include a bug report in our dataset only if there exists a corresponding commit in the collected commit history. 
%This provided us a total of 5,507 bug reports.
%We also include only such requests for which baseline queries cannot return their first relevant results within the Top-10 positions, \ie\ more effective search terms are required for those cases.
We first extract resolved bug reports (\ie\ marked as RESOLVED) from BugZilla and JIRA repositories, and then collect corresponding bug-fixing commits from GitHub version control histories of these eight systems.
Such approach was regularly adopted by the relevant literature \cite{refoqus,twkraft,sisman,saner2017masud}, and we also follow the same. 
In order to ensure a fair evaluation or validation, we discard the bug reports from our dataset for which no source code files (\eg\ Java classes) were changed or no relevant source files exist in the system snapshot collected for our study.
We also discard such bug reports that contain stack traces using appropriate regular expressions \cite{stacktrace}. 
They do not represent a typical change request (\ie\ mostly containing natural language texts) from the regular software users.

\textbf{Baseline Query Selection:} We select the \emph{title} of a bug report as the baseline query for our experiments, as was also selected by earlier studies \cite{refoqus,sisman,kevic}. 
However, we discard such queries that (\ie\ in verbatim titles) already return their first correct results within the Top-10 positions, \ie\ they possibly do not need query reformulation \cite{refoqus}.
Finally, we ended up with a collection of 1,675 baseline queries. 
We perform the same preprocessing steps as were done on the source documents (Section \ref{sec:preprocess}), on the queries before using them for code search in our experiments. 

\textbf{Goldset Development:}
%In GitHub and other version control systems, 
Developers often mention a Bug ID in the title of a commit when they fix the corresponding reported bug \cite{bugid}.  
%any commit that fixes a software bug generally mentions the corresponding Issue ID in the very title of the commit.
%We identify such commits from the commit history of each of our systems using suitable regular expressions, 
%and select them for our study \cite{bugid}. 
We collect the \emph{changeset} (\ie\ list of changed files) from each of our selected bug-fixing commits, 
and develop individual solution set (\ie\ \emph{goldset}) for each of the corresponding bug reports.  
Such solution sets are then used for the evaluation and validation of our suggested queries.
%Thus, for experiments, we collect not only the actual change requests from the reputed subject systems but also their solutions which were applied in practice by the developers \cite{specificity}.
%We use several utility commands such as \texttt{git, clone, rev-list} and \texttt{log} on GitHub console for collecting those information. 

\textbf{Replication:} All experimental data and relevant materials are hosted online \cite{acer} for replication or third party reuse.

\subsection{Corpus Indexing \& Source Code Search}\label{sec:codesearch}
%\textbf{Corpus Indexing \& Code Search Engine:}
Since we locate concept within project source, each of the source files is considered as an individual document of the corpus \cite{saha}. 
We apply the same preprocessing steps on the corpus documents as were done for query reformulation (\ie\ details in Section \ref{sec:preprocess}).
We remove punctuation marks and stop words from each document.
Then, we split the structured tokens, and keep both the original and the splitted tokens in the preprocessed documents.
We then apply \emph{Apache Lucene}, a \emph{Vector Space Model (VSM)} based popular search engine, to index all the documents and to search for relevant documents from the corpus for any given query. Such approaches and tools were widely adopted by earlier studies \cite{refoqus,flat3,saner2017masud,kevic}.

%Then we apply \emph{Apache Lucene Indexer} to index the corpus documents for each of our selected systems, and 
%
%as was also done by the earlier studies \cite{refoqus,saner2017masud}.
%It should be noted that we index both stemmed and non-stemmed version of our corpus, and conduct experiments with both of them.  
%%generally avoid stemming in the corpus indexing (rationale discussed in Section \ref{sec:preprocess}).
%
%\textbf{Code Search Engine:} We perform a re-enactment based evaluation on our reformulated queries. That is, as a developer would do, we attempt to find out whether the query can return the documents that were changed for addressing a change request or not. 
%We use a \emph{Vector Space Model (VSM)} based search engine-- \emph{Apache Lucene} \cite{refoqus}-- for our code search during evaluation and validation.
%Once a search is initiated using a query, the search engine prepares a short list by filtering out the irrelevant corpus documents using a \emph{Boolean Search Model}.
%Then it returns a ranked list of relevant documents by applying a \emph{TF-IDF} based scoring technique in association with \emph{VSM} \cite{saner2017masud}.
%As existing studies suggest \cite{antoniol, kevic, marcus}, we then consider Top-K results from the search engine for performance evaluation and validation of our suggested queries.

\subsection{Performance Metrics} \label{pmetrics}

\textbf{Query Effectiveness (QE):} It approximates the effort required to find out the first correct result for a query.
In other words, query effectiveness is defined as the rank of the first correct result returned by the query \cite{stacktrace}.
The lower the effectiveness score, the better the query is.

\textbf{Mean Reciprocal Rank (MRR):} Reciprocal rank is defined as the multiplicative inverse of query effectiveness measure.  
Mean Reciprocal Rank averages such measures for all the queries.
The higher the MRR value, the better the query is.

%\textbf{Mean Average Precision@K (MAP@K)}: Precision@K determines precision at the occurrence of every relevant result in the ranked list. Average Precision@K (AP@K) %averages the precision@K for all relevant results in the list for a search query. 
%Thus, Mean Average Precision@K (MAP@K) is calculated from the mean of average precision@K for all queries in the dataset.
% as follows:
%\begin{equation*}\label{eq:avep}
%\setlength\abovedisplayskip{0pt}
%\setlength\belowdisplayskip{0pt}
%AP@K=\frac{\sum_{i=1}^{M} P_{i}\times rel_i}{\left |S \right |},~~ MAP@K=\frac{\sum_{q\epsilon Q}APK(q)}{\left |Q\right |}
%\end{equation*}
%%\begin{equation*}
%%\setlength\abovedisplayskip{0pt}
%%\setlength\belowdisplayskip{0pt}
%%
%%\end{equation*}
%Here, $rel_{i}$ is the relevance function of $i^{th}$ result in the ranked list,  $P_{i}$ denotes the precision at $i^{th}$ result, and $M$ refers to number of total results. 
%$S$ is the solution set for a query, and $Q$ is the set of all queries.

\textbf{Top-K Accuracy:} It refers to the percentage of queries by which at least one correct result is returned within the Top-K results. 
The higher the metric value, the better the queries are.

%\textbf{Mean Recall (MR)}: \emph{Recall} denotes the fraction of the solution set that is retrieved for a search query. \emph{Mean Recall} averages such measures for all queries in the dataset.

\subsection{Evaluation of ACER and CodeRank}\label{sec:result}
We evaluate our technique using 1,675 baseline queries from eight subject systems and three performance metrics discussed above.
%We consider the \emph{title} from each request as a baseline query and employ re-enactment based evaluation as was also done by the earlier studies.  
We determine effectiveness and retrieval performance of our suggested reformulated queries, and then compare them with their baseline counterparts.
We also contrast our term weight with traditional term weights, and calibrate our technique using various configurations.
%Tables \ref{table:effectiveness}, \ref{table:baseline}, \ref{table:baselineret}, \ref{table:coderank} and Figures \ref{fig:topkeff} summarize our findings as follows:

\textbf{Answering RQ$\mathbf{_1}$--Effectiveness of ACER Queries:}
Table \ref{table:effectiveness} and \ref{table:baseline} show the effectiveness of ACER queries.
If our query returns the first correct result closer to the top position than the baseline query, then we consider that as \emph{query improvement}, and the vice versa as \emph{query worsening}. If both queries return their first correct results at the same position, we cosider that as \emph{query preserving}. 
From Table \ref{table:effectiveness}, we see that ACER can improve or preserve 97\% of the baseline queries (\ie\ about 71\% improvement and about 26\% preserving) while worsening the quality of only about 3\% of the queries.
All these statistics are highly promising according to the relevant literature \cite{refoqus,trconfig,saner2017masud}, \ie\ maximum 52\% improvement reported \cite{refoqus}, and they demonstrate the potential of our technique.
When individual systems are considered, our technique provides 63\%--82\% improvement across eight systems.
According to the quantile analysis in Table \ref{table:effectiveness}, 25\% of our queries return their first correct results within the Top-10 positions for all the systems except two (\ie\ Top-12 position for \texttt{log4j} and Top-21 position for \texttt{tomcat70}). 
Please note that only 6\% of the baseline queries return their correct results within the Top-10 positions (Table \ref{table:baselineret}). On the contrary, 25\% of our queries do so for six out of eight systems, which demonstrates the potential of our technique.
While query improvement ratios are significantly higher than the worsening ratios (\ie\ 28 times higher), it should be noted that our technique does not worsen any of the queries for two of the systems--\texttt{log4j} and \texttt{sling}. 

Table \ref{table:baseline} reports further effectiveness and the extent of actual rank improvements by our suggested queries.
%and ACER queries.
% extracted from method signatures and field signatures.
We see that reformulations from the method signatures improve the baseline queries significantly.
For example, they improve 59\% of the baseline queries while worsening 38\% of them. 
Reformulations from the field signatures are found relatively less effective.
However, ACER reduces the worsening ratio to as low as 2.51\%, and increases the improvement ratio up to 71\%, which are highly promising.  
More importantly, the mean rank differences (MRD) suggest that ACER \emph{elevates} first correct results in the ranked list by \textbf{81} positions on average for at least 71\% of the queries
while \emph{dropping} them for only 3\% of the queries by 104 positions.     
Such rank improvements are likely to reduce human efforts significantly during concept location.    
%However, our technique--ACER--not only captures the strength of the candidate reformulations but also overcomes their weaknesses, 
%and suggests the best reformulation for a query by employing query difficulty analysis and machine learning \cite{qperf}.
%Thus, ACER reduces the worsening ratio to as low as 2.51\%. Simultaneously, it leads to a query improvement ratio of 71\% which is highly promising.  

\begin{table}[!t]
\centering
\caption{Effectiveness of ACER Variants against Baseline Queries}\label{table:baseline}
\vspace{-.2cm}
\resizebox{3.5in}{!}{%
\begin{threeparttable}
\begin{tabular}{l|c|c|c|c}
%\hline
%\multicolumn{4}{c||}{\textbf{Java Systems}} & \multicolumn{2}{c}{\textbf{C Systems}}\\
\hline
\textbf{Query Pairs} & \textbf{Improved} (MRD) & \textbf{Worsened} (MRD)  & \textbf{p-value} & \textbf{Preserved} \\
\hline
\hline
ACER$_{msig}$ vs. Baseline & \textbf{58.93}\% (-61) &  37.99\% (+131) & \textbf{*0.007} & \textbf{3.08}\% \\
\hline
ACER$_{fsig}$ vs. Baseline & 52.51\% (-51) &  44.57\% (+151) & 0.063 & \textbf{2.91}\% \\
\hline
ACER$_{comb}$ vs. Baseline & \textbf{58.62}\% (-51) & 38.19\% (+136) & \textbf{*0.018} & \textbf{3.20}\% \\
\hline
\textbf{ACER vs. Baseline} & \textbf{71.05\% (-81)}  &  2.51\% (+104) & \textbf{*$<$0.001} & \textbf{26.44}\% \\
\hline
%ACER$_{best}$+$Q_{pp}$ & \textbf{59.40}\% (-41) &  38.32\% (89)&  \textbf{*0.002}  & \textbf{2.28}\%  \\
%\hhline{~~~~~}
%vs. Baseline & & & & \\
%\hline
%ACER (ACER$_{msig}$+$Q_{pp}$) & \textbf{60.77}\% (-51)  &  35.72\% (89) & \textbf{*0.001} & \textbf{3.51}\% \\
%\hhline{~~~~~}
%vs. Baseline & & & & \\
%\hline
\end{tabular}
%\begin{tablenotes}
%\item [1] No. of example pairs for which relative quality evaluation matches with that of StackOverflow
%\item [2] \% of agreement, \item [3] \% of disagreement
\centering
\textbf{*} = Statistically significant difference between improvement and worsening, \textbf{MRD} = Mean Rank Difference between ACER and baseline queries 
 %\end{tablenotes}
\end{threeparttable}
%\vspace{-.6cm}
}
\vspace{-.7cm}
\end{table}

\textbf{Retrieval Performance of ACER Queries:} 
%\citet{qperf} argue that unlike information retrieval, concept location emphasizes on finding the first correct result close to the top of the list rather than finding all correct results.
%That is, precision or recall might not be the most appropriate metric for determining the retrieval performance of the queries during concept location. 
Table \ref{table:baselineret} reports the comparison of retrieval performance between our queries and baseline queries.
Given that most of our selected queries are difficult (\ie\ no correct results within the Top-10 positions \cite{refoqus}), the baseline queries retrieve at least one correct result within the Top-100 positions for 56\% of the cases. 
However, our reformulations improve this ratio to about 64\%, and the improvement is statistically significant (\ie\ \emph{paired t-test, p-value=0.010$<$0.05}, \emph{Cohen's D=0.68 (moderate)}). 
Similar scenarios are observed with mean reciprocal rank as well. 

Thus, to answer \textbf{RQ$\mathbf{_1}$}, the reformulation of ACER improves the baseline queries significantly both in terms of query effectiveness and retrieval performance.  
ACER improves 71\% of the baseline queries with 64\% Top-100 retrieval accuracy.

%All these findings suggest the high potential of our technique for effective query reformulation. 

\textbf{Answering RQ$\mathbf{_2}$--CodeRank vs. Traditional Term Weighting Methods:}
Table \ref{table:coderank} shows the comparative analysis between CodeRank and two traditional term weights--TF and TF-IDF-- which are widely used in the text retrieval contexts \cite{rocchio,kevic,qsurvey}.
While TF estimates the importance of a term based on its occurrences within a document, 
TF-IDF additionally captures the global occurrences of the term across all the documents of the corpus \cite{tfidf}.
On the contrary, CodeRank employs a graph-based scoring mechanism that determines the importance of a term based on its co-occurrences with other important terms within a certain context.  
From Table \ref{table:coderank}, we see that CodeRank performs significantly better than both TF (\ie\ \emph{paired t-test, p-value=0.005$<$0.05}) and TF-IDF (\ie\ \emph{p-value$<$0.001}) in identifying important search terms from source code, especially from the method signatures.
Considering the whole source code rather than signatures improves the performance of both TF (\ie\ 56\% query improvement) and TF-IDF (\ie\ 52\% query improvement). However, our term weight--CodeRank--is still better alone (\ie\ 59\%), 
and improves significantly higher (\ie\ \emph{p-value=1.717e-06}) fraction (\ie\ 71\%) of the baseline queries when employed with our proposed reformulation algorithm--ACER.

\begin{figure}[!t]
\centering
\includegraphics[width=3.45in]{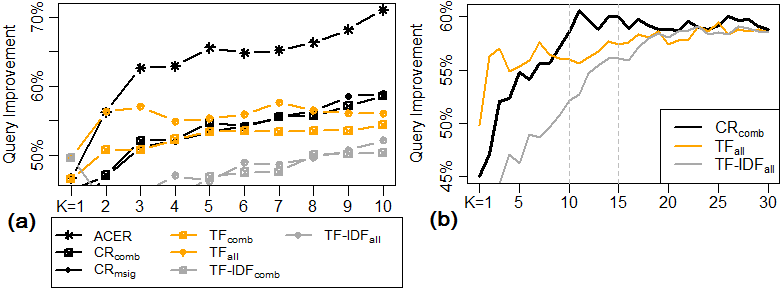}
\vspace{-.7cm}
\centering
\caption{Comparison of query improvement between CodeRank and traditional term weights for (a) Top-10 and (b) Top-30 reformulated query terms}
\vspace{.1cm}
\label{fig:topkeff}
\end{figure}

Fig. \ref{fig:topkeff} shows how \emph{CodeRank} and traditional term weights perform in reformulating the baseline queries with their (a) Top-10 and (b) Top-30 terms.
We see that TF reaches its peak performance pretty quickly (\ie\ $K=3$), and then shows a stationary or irregular behaviour.
That means, TF identifies frequent terms for query reformulation, and few of them (\eg\ Top-3) could be highly effective.
On the contrary, our method--CodeRank-- demonstrates a gradual improvement in the performance up to Top-12 terms (\ie\ $K$=12, Fig. \ref{fig:topkeff}-(b)), and 
crosses the performance peak of TF with a large margin (\ie\ \emph{paired t-test, p-value=0.004$<$0.05, Cohen's D=3.77$>$1.00 (large))}, for $K$=10 to $K$=15). 
CodeRank emphasizes on the votes from other important terms (\ie\ by leveraging co-occurrences) for determining weight of a term, and as demonstrated in Fig. \ref{fig:topkeff}, this weight is found to be more reliable than TF.
TF-IDF is found relatively less effective according to our investigation.

\begin{table}[!t]
\centering
\caption{Comparison of ACER's Retrieval Performance with Baseline Queries}\label{table:baselineret}
\vspace{-.2cm}
\resizebox{3.3in}{!}{%
\begin{threeparttable}
\begin{tabular}{l|l|c|c|c|c}
%\hline
%\multicolumn{4}{c||}{\textbf{Java Systems}} & \multicolumn{2}{c}{\textbf{C Systems}}\\
\hline
\textbf{Query} & \textbf{Metric} & \textbf{Top-10} & \textbf{Top-20} & \textbf{Top-50} & \textbf{Top-100} \\
\hline
\hline
\multirow{2}{*}{Baseline} & Top-K Accuracy & 5.78\% & 18.91\% & 41.09\% & 56.30\% \\
\hhline{~-----}
& MRR@K & 0.01 & 0.02 & 0.03 & 0.03  \\
%\hhline{~-----}
%& MAP@K & 1.15\% & 2.08\% & 2.84\% & 3.03\% \\
\hline
%\multirow{3}{*}{Baseline$_{stem}$}  & Top-K Accuracy & \textbf{14.33}\% & 26.08\% & 45.72\% & 58.67\%  \\
%\hhline{~-----}
%& MRR@K & 0.05 & 0.06  & 0.06 & 0.07 \\
%\hhline{~-----}
%& MAP@K & 5.10\%  & 5.86\% & 6.15\% & 6.09\% \\
%\hline
\multirow{2}{*}{ACER$_{msig}$}& Top-K Accuracy & 10.45\% & 21.48\% & 38.12\% &  51.31\% \\
\hhline{~-----}
& MRR@K & 0.02 & 0.03 & 0.04 & 0.04\\
%\hhline{~-----}
%& MAP@K & 2.42\% & 3.20\%  & 3.48\%  & 3.43\%\\
\hline
\multirow{2}{*}{ACER$_{fsig}$}& Top-K Accuracy & 7.77\% & 17.40\% & 36.25\% &  47.23\% \\
\hhline{~-----}
& MRR@K & 0.02 & 0.03 & 0.03 & 0.03\\
\hline
\multirow{2}{*}{ACER$_{comb}$}& Top-K Accuracy & 8.68\% & 20.78\% & 36.87\% &  51.75\% \\
\hhline{~-----}
& MRR@K & 0.02 & 0.03 & 0.03 & 0.04\\
%\hhline{~-----}
%& MAP@K & 2.09\% & 2.87\%  & 3.15\%  & 3.16\%\\
\hline
\multirow{2}{*}{ACER}& Top-K Accuracy &  \textbf{*14.72}\% & \textbf{*31.22}\% & \textbf{*49.89}\% &  \textbf{*63.89}\% \\
\hhline{~-----}
& MRR@K & 0.04 & 0.05 & \textbf{0.06} & \textbf{0.06}\\
%\hhline{~-----}
%& MAP@K & 3.54\% & 4.58\%  & \textbf{4.83}\%  & \textbf{4.66}\%\\
\hline
\end{tabular}
%\begin{tablenotes}
%\item [1] No. of example pairs for which relative quality evaluation matches with that of StackOverflow
%\item [2] \% of agreement, \item [3] \% of disagreement
\centering
\textbf{*} = Statistically significant difference between ACER and baseline
 %\end{tablenotes}
\end{threeparttable}
%\vspace{-.2cm}
}
%\vspace{-.3cm}
\end{table}

Thus, to answer \textbf{RQ$\mathbf{_2}$}, CodeRank performs significantly better than traditional methods in identifying effective terms for query reformulation from the source code.

\begin{table}[!t]
\centering
\caption{Comparison between CodeRank and Traditional Term Weights}\label{table:coderank}
\vspace{-.2cm}
\resizebox{3.5in}{!}{%
\begin{threeparttable}
\begin{tabular}{l|c|c|c}
%\hline
%\multicolumn{4}{c||}{\textbf{Java Systems}} & \multicolumn{2}{c}{\textbf{C Systems}}\\
\hline
\textbf{Query Pairs} & \textbf{Improved} & \textbf{Worsened} & \textbf{Preserved} \\
\hline
\hline
ACER$_{msig}$ vs. TF$_{msig}$   & \textbf{*58.93}\% / 53.40\% &  \textbf{*37.99}\% / 44.60\%  & 3.08\% / 2.00\% \\
\hline
ACER$_{fsig}$ vs. TF$_{fsig}$ & 52.51\% / 51.57\% & 44.57\% / 46.85\%  & 2.91\% / 1.57\% \\
\hline
ACER$_{comb}$ vs. TF$_{comb}$ & \textbf{*58.62}\% / 54.34\% & \textbf{*38.19}\% / 44.11\% & 3.20\% / 1.54\% \\
\hline
ACER vs. TF$_{all}$ & \textbf{*71.05}\% / 56.01\% &  \textbf{*2.51}\% / 41.44\%  & \textbf{*26.44}\% / 2.55\% \\
\hline
\hline
ACER$_{msig}$ vs.    & \multirow{2}{*}{\textbf{*58.93}\% / 45.55\%} &  \multirow{2}{*}{\textbf{*37.99}\% / 49.88\%}  & \multirow{2}{*}{3.08\% / 4.57\%} \\
TF-IDF$_{msig}$ & &  &  \\
\hline
ACER$_{fsig}$ vs.  & \multirow{2}{*}{52.51\% / 51.06\%} & \multirow{2}{*}{44.57\% / 46.77\%}  & \multirow{2}{*}{2.91\% / 2.17\%} \\
TF-IDF$_{fsig}$ & &  &  \\
\hline
ACER$_{comb}$ vs.  & \multirow{2}{*}{\textbf{*58.62}\% / 50.35\%} & \multirow{2}{*}{\textbf{*38.19}\% / 47.25\%} & \multirow{2}{*}{3.20\% / 2.40\%} \\
TF-IDF$_{comb}$ & &  &  \\
\hline
ACER vs.  & \multirow{2}{*}{\textbf{*71.05}\% / 52.17\%} &  \multirow{2}{*}{\textbf{*2.51}\% / 45.13\%} &  \multirow{2}{*}{\textbf{*26.44}\% / 2.70\%} \\
TF-IDF$_{all}$ & &  &  \\
\hline
\end{tabular}
%\begin{tablenotes}
%\item [1] No. of example pairs for which relative quality evaluation matches with that of StackOverflow
%\item [2] \% of agreement, \item [3] \% of disagreement
\centering
\textbf{*} = Statistically significant difference between ACER measures and their counterparts
 %\end{tablenotes}
\end{threeparttable}
%\vspace{-.2cm}
}
\vspace{-.4cm}
\end{table}

\begin{figure}[!t]
\centering
\includegraphics[width=3.5in]{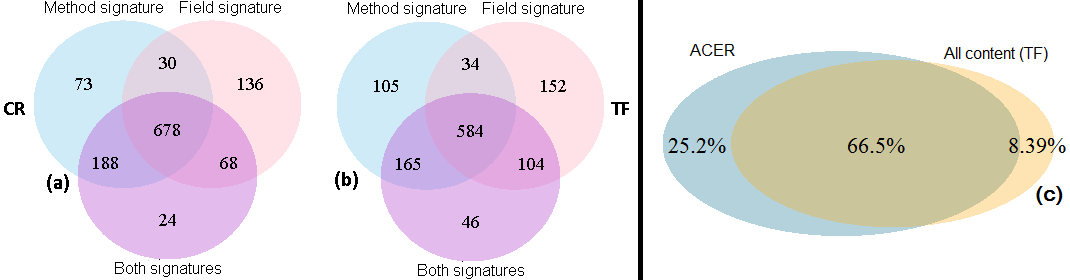}
\vspace{-.6cm}
\caption{Improved queries by reformulation from method signatures and field signatures using (a) CodeRank (CR) and (b) Term Frequency (TF). (c) ACER vs. TF (all content)}
%\vspace{-.2cm}
\label{fig:vennsigs}
\end{figure}

%\begin{figure}[!t]
%\centering
%\includegraphics[width=1.5in]{acer-tf-venn}
%\vspace{-.3cm}
%\caption{Improved queries by (a) reformulation of ACER using method signatures and field signatures, and (b) reformulation of traditional method (TF) using all source content}
%\vspace{-.7cm}
%\label{fig:venncrcf}
%\end{figure}

\textbf{Answering RQ$\mathbf{_3}$--Do Document Structures Matter?} While most of the earlier reformulation techniques miss or ignore the structural aspect of a source document, we consider such aspect as an important paradigm of our technique.
We consider a source document as a collection of structured entities (\eg\ signatures, methods, fields) \cite{scam2014masud} rather than a regular text document. Thus, we make use of method signatures and field signatures rather than the whole source code for query reformulation given that they are likely to contain more salient terms and less noise \cite{hillicse09}. 
Fig. \ref{fig:vennsigs} demonstrates how incorporation of document structures 
into a technique could be useful for query reformulations.
We see that reformulations using method signatures and field signatures improve two different sets of baseline queries, and this happens with both term weighting methods--(a) CodeRank and (b) TF. While these sets share about half of the queries (49\%--57\%), reformulations based on each signature type also improve a significant amount (\ie\ 19\% (73+136+24) -- 25\% (105+152+46)) of unique baseline queries.
In Fig. \ref{fig:vennsigs}-(c), when these signatures (\ie\ along with ACER) are contrasted with the whole source code (\ie\ along with TF), we even found that the signature-based reformulations outperform the whole code-based reformulations by a large margin (\ie\ (25.2\%--8.39\%) $\approx$ 17\% more query improvement).
%improve significantly higher fraction (\ie\ 17\%) of the unique queries. 
That is, the use of the whole source code introduces additional noise, and diminishes the strength or salience of the individual structures (\ie\ signatures). Most of the existing methods \cite{refoqus,gayg,ase2016masud} suffer from this limitation.
On the contrary, our technique ACER exploits document structures (\ie\ signatures), and carefully chooses the best among all the candidate reformulations derived from such structures using query quality analysis and machine learning.

Thus, to answer \textbf{RQ$\mathbf{_3}$}, document structures improve the suggestion of query reformulation terms from the source code.

\begin{table}[!t]
\centering
\caption{Impact of Stemming on Query Effectiveness}\label{table:stemming}
\vspace{-.2cm}
\resizebox{3.5in}{!}{%
\begin{threeparttable}
\begin{tabular}{l|l|c|c|c}
%\hline
%\multicolumn{4}{c||}{\textbf{Java Systems}} & \multicolumn{2}{c}{\textbf{C Systems}}\\
\hline
\textbf{Source} & \textbf{Query} & \textbf{Improved} (MRD) & \textbf{Worsened} (MRD)  & \textbf{Preserved} \\
\hline
\hline
Method & ACER$_{msig,stem}$ & \textbf{52.66}\% (-58) &  44.73\% (+127)  & 2.61\% \\
\hhline{~----}
signature & ACER$_{msig}$ & *\textbf{58.93}\% (-61) &  *\textbf{37.99}\% (+131)  & 3.08\% \\
\hline
Field & ACER$_{fsig,stem}$ & 48.14\% (-53) &  47.47\% (+151) & 4.39\% \\
\hhline{~----}
signature & ACER$_{fsig}$ & 52.51\% (-51) &  44.57\% (+151)  & 2.91\% \\
\hline
Both & ACER$_{comb,stem}$ & 52.68\% (-57) & 44.38\% (+128) & 2.94\% \\
\hhline{~----}
signatures & ACER$_{comb}$ & *58.62\% (-51) & *38.19\% (+136)  & 3.20\% \\
\hline
Both & ACER$_{stem}$ & \textbf{68.11}\% (-78)  &  5.37\% (+67) & \textbf{26.51}\% \\
\hhline{~----}
signatures & ACER & \textbf{71.05}\% (-81)  &  *\textbf{2.51}\% (+104)  & \textbf{26.44}\% \\
\hline
%ACER$_{best}$+$Q_{pp}$ & \textbf{59.40}\% (-41) &  38.32\% (89)&  \textbf{*0.002}  & \textbf{2.28}\%  \\
%\hhline{~~~~~}
%vs. Baseline & & & & \\
%\hline
%ACER (ACER$_{msig}$+$Q_{pp}$) & \textbf{60.77}\% (-51)  &  35.72\% (89) & \textbf{*0.001} & \textbf{3.51}\% \\
%\hhline{~~~~~}
%vs. Baseline & & & & \\
%\hline
\end{tabular}
%\begin{tablenotes}
%\item [1] No. of example pairs for which relative quality evaluation matches with that of StackOverflow
%\item [2] \% of agreement, \item [3] \% of disagreement
\centering
\textbf{*} = Statistically significant difference between two measures from the same signature, \textbf{MRD} = Mean Rank Difference between ACER and baseline queries 
 %\end{tablenotes}
\end{threeparttable}
%\vspace{-.2cm}
}
\vspace{-.7cm}
\end{table}

\begin{table*}[!t]
\centering
\caption{Comparison of Query Effectiveness with Existing Techniques}\label{table:comparison}
\vspace{-.2cm}
\resizebox{7.2in}{!}{%
\begin{threeparttable}
\begin{tabular}{l|c|c|c|c|c|c|c|c||c|c|c|c|c|c|c||c}
\hline
\multirow{2}{*}{\textbf{Technique}} & \multirow{2}{*}{\textbf{\#Queries}} &\multicolumn{7}{c||}{\textbf{Improvement}} & \multicolumn{7}{c||}{\textbf{Worsening}} & \textbf{Preserving}   \\
%\hhline{~~---------------}
%\textbf{Metric}  & \multicolumn{2}{c|}{\textbf{Class Name}}  & \multicolumn{2}{c|}{\textbf{Method Name}} & \multicolumn{2}{c|}{\textbf{Variable Name}} & \multicolumn{2}{c||}{\textbf{All Combined}}  & \multicolumn{2}{c|}{\textbf{Class Name}}  & \multicolumn{2}{c|}{\textbf{Method Name}} & \multicolumn{2}{c|}{\textbf{Variable Name}} & \multicolumn{2}{c}{\textbf{All Combined}}\\
\hhline{~~---------------}
& & \#Improved & Mean & Q1 & Q2 & Q3 & Min. & Max. & \#Worsened & Mean & Q1 & Q2 & Q3 & Min. & Max. & \#Preserved\\
\hline
\hline
\citet{hillicse09} & 1,675 & 631 (37.67\%) & 157 & 18 & 48 &  161 & 1 & 2,264 &  760 (45.37\%) & 261 & 54 & 119 & 300 & 4 & 4,819 & 284 (16.96\%) \\
\hline
\citet{rocchio}  & 1,675 & 895 (53.43\%) & 219 & 15 & 49 & 188 & 1 & 4,609 &	 739 (44.11\%) & 333 & 65 & 170 & 429 & 3 & 3,489 & 41 (2.45\%)\\
\hline
RSV \cite{qsurvey} & 1,675 & \textbf{914 (54.57\%)} & 216 & 15 &  52 & 195 & 1 & 4,611 & 723 (43.16\%) & 307 & 63 & 160 & 415 & 7 & 3,387 & 38 (2.27\%)\\
\hline
\citet{sisman} & 1,675 & 759 (45.31\%) & 207 & 17 &  61 & 213 & 1 & 3,707 &	642 (38.33\%) & 273 & 59 & 147 & 345 & 8 & 2,545 & 274 (16.36\%)\\
\hline
\textbf{Refoqus} \cite{refoqus} & 1,675 & \textbf{895 (53.43\%)} & 217 & 15 & 51 & 188 & 1 & 4,609 &  737 (44.00\%) & 332 & 65 & 170 & 429 & 3 & 3,489 & 43 (2.57\%)\\
\hline
\textbf{Refoqus}$_{sampled}$ \cite{refoqus} & 1,675 & \textbf{1,154 (68.90\%)} & 156 & 11 & 33 & 141 & 1 & 4,609 & \textbf{487 (29.07\%)} & 325 & 63 & 166 & 406 & 6 & 3,489 & \textbf{34 (2.03\%)}\\
\hline
\hline
ACER$_{msig}$  & 1,675 & 969 \textbf{(57.85\%)} & 208 & 14 & 49 & 192 & 1 & 3,649 &  662 (39.52\%) & 272 & 52 & 139 & 341 & 2 & 4,825 & 44 (2.63\%)\\
\hline
ACER$_{comb}$  & 1,675 & 958 (57.19\%) & 216 & 15 & 49 & 194 & 1 & 4,117 & 674 (40.24\%) & 275 & 52 & 139 & 336 & 4 & 3,360 & 43 (2.57\%)\\
\hline
%\textbf{ACER$_{stem}$} & 1,675 & \textbf{*1,145 (68.36\%)} & 158 & \textbf{11} & 38 & 148 & 1 & 3,233 & \textbf{*96 (5.73\%)} & 201 & 28 & 71 & 210 & 3 & 1,669 & \textbf{*432 (25.79\%)} \\ 

%\hline
\textbf{ACER} & 1,675 & \textbf{*1,169 (69.79\%)} & 156 & \textbf{11} &  35 & 130 & 1 & 3,162 & \textbf{*57 (3.40\%)} & 260 & 53 & 140 & 375 & 13 & 1,369 & \textbf{*449 (26.81\%)}\\
\hline
\textbf{Baseline} & 1,675 & - & 227 & 32 & 88 & 258 & 3 & 4,787 & - & 113 & 24 & 49 & 162 & 1 & 718 & - \\ 
\hline
\textbf{ACER$_{ext}$} & 1,755 & \textbf{*1,192 (67.92\%)} & 149 & \textbf{10} &  34 & 124 & 1 & 3,162 & \textbf{*48 (2.74\%)} & 301 & 50 & 145 & 327 & 13 & 1,782 & \textbf{*515 (29.34\%)}\\
%& Total=1,675 & \textbf{Avg = 71.05\%}  & &  &  &  &  &  & Avg = 2.51\%  &  &  &  &  &  &  & \textbf{Avg = 26.44\%}    \\
\hline
%\textbf{Recall} & 36.91\% & 59.18\% & 38.40\% & 59.25\% & 38.36\% & 59.85\% & 42.70\% & \textbf{62.83}\% & 37.50\% & 55.34\% & 39.85\% & 58.20\% & 39.78\% & 57.91\% & 43.03\% & \textbf{61.54}\%\\
%\hline
\end{tabular}
%\begin{tablenotes}
%\item [1] No. of example pairs for which relative quality evaluation matches with that of StackOverflow
%\item [2] \% of agreement, \item [3] \% of disagreement
 %\end{tablenotes}
\centering
%\textbf{jdt.core} = \texttt{eclipse.jdt.core}, \textbf{jdt.debug} = \texttt{eclipse.jdt.debug}, \textbf{jdt.ui} = \texttt{eclipse.jdt.ui}, \textbf{pde.ui} = \texttt{eclipse.pde.ui}, 
\textbf{Mean} = Mean rank of first correct results returned by the queries, \textbf{Q$_i$}=  $i^{th}$ quartile of all ranks considered, \textbf{*} = Statistically significant difference between ACER measures and their counterparts
\end{threeparttable}
%\vspace{-.2cm}
}
\vspace{-.6cm}
\end{table*}

%\begin{figure}[!t]
%\centering
%\includegraphics[width=1.5in]{rf-topk}
%\vspace{-.3cm}
%\caption{Query effectiveness for Top-K relevance feedback documents}
%\vspace{-.6cm}
%\label{fig:rftopk}
%\end{figure}

\begin{figure}[!t]
\centering
\includegraphics[width=3.3in]{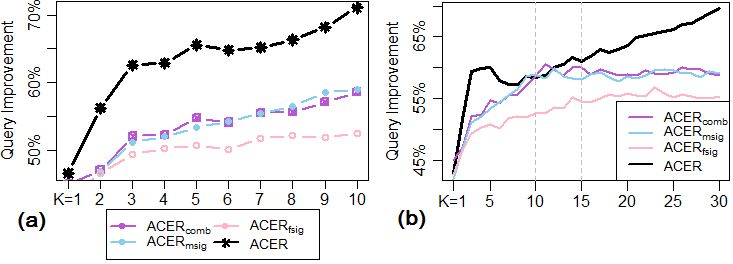}
\vspace{-.4cm}
\centering
\caption{Effectiveness of ACER queries for (a) Top-10 and (b) Top-30 reformulated terms}
\vspace{-.5cm}
\label{fig:topkqeff}
\end{figure}

\textbf{Answering RQ$\mathbf{_4}$-- Impact of Stemming, Query Length, and Relevance Feedback:}
From Table \ref{table:stemming}, we see that stemming generally degrades the effectiveness of our reformulated queries.  
Similar findings were also reported by earlier studies \cite{kevic,saha}.
Fig. \ref{fig:topkqeff} shows how (a) Top-10 and (b) Top-30 reformulation terms improve the baseline queries.
We see that our reformulations perform the best (\ie\ about 60\% query improvement) with Top-10 to 15 search terms collected from each signature type.
However, when query quality analysis \cite{qperf} is employed, our technique--ACER--can improve 71\% of the baseline queries with only Top-10 reformulation terms.
We also repeat the same investigation with Top-30 terms, and achieved the same top performance (\ie\ Fig. \ref{fig:topkqeff}-(b)). Thus, our choice of returning Top-10 reformulation terms is justified.
We also investigate how the size of pseudo-relevance feedback influences our performance, and experimented with Top-30 documents.
%From Fig. \ref{fig:rftopk}, 
We found that reformulations for ACER reach the performance peak 
when Top-10 to 15 feedback source documents (\ie\ returned by the baseline queries) are analyzed for candidate terms.
% as a pseudo-relevance feedback. 
This possibly justifies our choice of 
%Such findings also possibly justify our choice of considering the
Top-10 documents as the pseudo-relevance feedback.

Thus, to answer \textbf{RQ$\mathbf{_4}$}, stemming degrades the query effectiveness of ACER. Reformulation size and relevance feedback size gradually improve the performance of ACER as long as they are below a certain threshold (\ie\ $K=15$).

%\begin{figure}[!t]
%\centering
%\includegraphics[width=1.85in]{compare-topk}
%\vspace{-.3cm}
%\caption{Comparison of Top-K accuracy with existing techniques}
%\vspace{-.8cm}
%\label{fig:compare-topk}
%\end{figure}

\subsection{Comparison with Existing Approaches}\label{sec:comparison}
\begin{figure}[!t]
\centering
\includegraphics[width=3.5in]{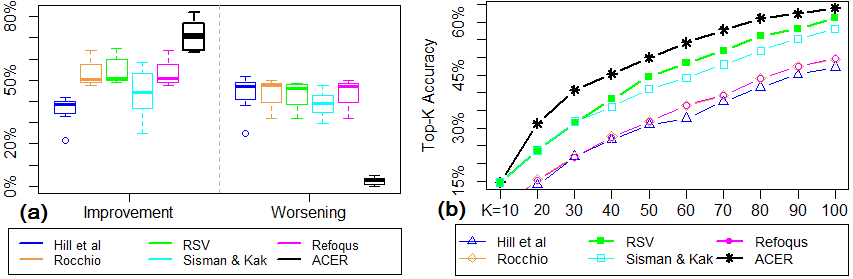}
\vspace{-.7cm}
\caption{Comparison of (a) query effectiveness, and (b) retrieval performance}
\vspace{-.6cm}
\label{fig:compare-box}
\end{figure}

\textbf{Answering RQ$\mathbf{_5}$:} While the empirical evaluation in terms of performance metrics above clearly demonstrates the promising aspects of our query reformulation technique, we still compare with five closely-related existing approaches \cite{hillicse09,rocchio,qsurvey,sisman,refoqus}.
\citet{hillicse09} suggest relevant phrases from method signatures and field signatures as query reformulations. While \citet{sisman} focus on term co-occurrences with query keywords, \citet{rocchio} and RSV \cite{qsurvey} apply TF-IDF based term weights for choosing query reformulation terms. Refoqus \cite{refoqus} is closely related to ours and is reported to perform better than RSV and other earlier approaches, which probably makes it the state-of-the-art for our research problem.
We replicate each of \citeauthor{hillicse09}, \citeauthor{rocchio}, RSV, \citeauthor{sisman}, and Refoqus in our working environment by carefully following their algorithms, equations and methodologies given that their implementations are not publicly available.
In the case of Refoqus, we implement 27 metrics (20 pre-retrieval \cite{qperf} and 7 post-retrieval \cite{refoqus}) that estimate query difficulty. 
%One metric was omitted due to redundancy.
We develop a machine learning model using CART algorithm (\ie\ as used by them) and 10-fold cross validation. Then, we use the model to return the best reformulation out of four candidates of Refoqus-- \emph{query reduction, Dice expansion, Rocchio's expansion} and \emph{RSV expansion}--for each baseline query.
Table \ref{table:comparison} and Fig. \ref{fig:compare-box} summarize our comparative analyses.

From Table \ref{table:comparison}, we see that RSV and Refoqus perform better than the other existing approaches.
They improve about 55\% and about 53\% of the baseline queries respectively. Such ratios are also pretty close to the originally reported performances by \citeauthor{refoqus} on a different dataset, which possibly validates the correctness of our implementation.
While 55\% query improvement is the maximum performance provided by any of the existing approaches, our technique--ACER--improves about 70\% of the baseline queries (\ie\ 1\% difference between Table \ref{table:baseline} and Table \ref{table:comparison} due to rounding error) which is significantly higher, \ie\ \emph{paired t-test, p-value=6.663e-06$<$0.05, Cohen's D=2.43$>$1.00 (large)}. Refoqus adopts a similar methodology like ours.
%, and chooses the best reformulation candidate using machine learning. 
Unfortunately, the approach is limited due to possibly the low performance of its  candidate reformulations. 
%As shown in Table \ref{table:comparison}, our candidate reformulations derived from the two signatures (\ie\ structural aspects) and CodeRank (\ie\ term weighting method) alone are more effective than their candidates and queries from the remaining approaches.
One might argue about the data resampling step (\ie\ Step 9, Fig. \ref{fig:sysdiag}) of ACER for the high performance.
However, we also apply data resampling to Refoqus using the same settings as ours for further investigation. We see that Refoqus$_{sampled}$ has a similar improvement ratio like ours, but it still worsens a significant amount of queries, 29\%, compared to our 3.40\%.
Thus, our technique still performs better than Refoqus in the equal settings.
%In short, Refoqus chooses the best among a set of mediocre candidates whereas our technique suggests the best from among a set of already good candidates (\eg\ ACER$_{msig}$) for query reformulation, which might have possibly made the difference.
%As shown in Table \ref{table:comparison}, our technique also performs significantly better than all the counterparts when worsening ratios and preserving ratios or any granular level statistics (\eg\ quantiles, mean ranks) are considered.
Our quantile measures and mean ranks are more promising than those from the baseline or competing methods as reported in Table \ref{table:comparison}. Table \textbf{\ref{table:baseline}} and \textbf{RQ$\mathbf{_1}$} also suggest that our queries have high potential for reducing human efforts. We also experiment with an extended dataset (\ie\ 1,755=1,675 + 8x10) containing 80 very good queries.
As reported in Table \ref{table:comparison}, ACER$_{ext}$ mostly preserves the good quality queries rather than worsening, which also demonstrates its high potential.

Fig. \ref{fig:compare-box}-(a) shows the box plots of query improvement and query worsening ratios by all the techniques under study. We see that ACER outperforms the existing techniques including the state-of-the-art \cite{refoqus} by a large margin.
Our median improvement ratio is about 75\%, which is higher than even the maximum improvement ratios of the counterparts, which demonstrates the promising aspect of ACER. Fig. \ref{fig:compare-box}-(b) shows the Top-K accuracy of the query reformulation techniques.
We see that our accuracy is relatively higher than that of each of the existing approaches across various Top-K (\ie\ 10--100) values.
The best performing existing method is RSV. However, our performance is significantly higher than that of RSV for various K values according to statistical significance tests (\ie\ \emph{paired t-test, p-value=0.0001$<$0.05, Cohen's D=0.34}). 

Thus, to answer \textbf{RQ$\mathbf{_5}$}, our technique outperforms the state-of-the-art techniques in terms of reformulation query effectiveness, and performs significantly better than each of the existing techniques in terms of document retrieval accuracy.

\section{Threats to Validity}\label{sec:threats}
Threats to \emph{internal validity} relate to experimental errors and biases \cite{wordsim}. 
Although CodeRank and document structures play a major role, the data resampling step (Section \ref{sec:bestref}, Step 9, Fig. \ref{fig:sysdiag}) has a significant role behind the high performance of our technique.
Unfortunately, to the best of our knowledge, Refoqus \cite{refoqus} does not have such a step.
Thus, the performance comparison might look like a bit unfair. Besides, models based on data resampling are sometimes criticized for intrinsic biases \cite{rscritic}.
However, we apply data resampling to Refoqus as well (\ie\ Refoqus$_{sampled}$), and demonstrate that our technique still performs better in terms of worsening ratio.
%However, we resampled the data 50 times, developed model for each sample with cross-validation, and average their predictions to make our decision on the best candidate.

Threats to \emph{external validity} relate to the generalization of the obtained results \cite{refoqus}.
All of our subject systems are Java-based.
% and they are from either \emph{Eclipse} or \emph{Apache}. 
So, there might be different results with systems from other programming languages.
However, we experimented with eight different systems with promising performance, and the comparison with the state-of-the-art techniques demonstrates the superiority of our approach. 

%However, our technique is compared with five closely related approaches, and ACER clearly performs better than each of them.
%However, our model might perform differently if it is trained using subject systems from unrelated domains or platforms.

%\vspace{-.2cm}
\section{Related Work}\label{sec:related}
There exist a number of studies in the literature that reformulate a given query for concept location in the context of software change tasks.
Existing studies apply relevance feedback from developers \cite{gayg}, pseudo-relevance feedback from IR tools \cite{refoqus}, partial phrasal matching \cite{hillicse09,hilltool}, and machine learning  \cite{refoqus,trconfig} to query reformulation. They also make use of context of query terms from source code \cite{ccmapping,infer,ase2016masud,sisman}, text retrieval configuration \cite{refoqus,trconfig}, and quality of queries \cite{specificity,qperf} in suggesting the reformulated queries.
\citet{hillicse09} consider the presence of query terms in the method or field signatures as an indicator of their relevance, and suggest natural language phrases from them as reformulated queries.
\citet{sisman} choose such terms for query reformulation that frequently co-occur with query terms within a fixed size of window
in the code. \citet{rocchio} and RSV \cite{qsurvey} determine importance of a term using TF-IDF based metrics.
\citet{refoqus} identify the best of four reformulation candidates for any given query using a machine learning model with 28 metrics. 
All these five studies are highly relevant to ours, and we directly compare with them using experiments. Readers are referred to Section \ref{sec:comparison} for comparison details.

%There exist studies \cite{saner2015masud,saner2017masud,dblp} that also adopt graph-based method for term weighting.
%%Please note that our work is fundamentally different from \citet{saner2017masud} which also adopts \emph{graph-based} term weighting for concept location.
%However, \citeauthor{saner2015masud} \cite{saner2015masud,saner2017masud} take a change request as input, and suggest \emph{initial queries} simply by using TextRank on the \emph{request texts}.
%On the contrary, we take initial queries as input,  and \emph{reformulate} them not only by employing our term weighting method--CodeRank--for \emph{source code}, but also by applying source document structures, query quality analysis and machine learning \cite{qperf}.
%Since they solve a different research problem and apply a different information source (\ie\ request texts) than ours, a direct comparison with them might not be applicable. Given that reformulation is performed on the initial queries, our technique can potentially complement theirs.
%%Furthermore, source code has different structures, contexts and often different semantics than natural language texts \cite{semantictool}. We thus face and overcome a different set of technical challenges (\eg\ source term graph development, reformulation candidate selection), which makes our work novel. 

Other related studies \cite{saner2015masud,saner2017masud,dblp} explore graph-based methods for term weighting. \citeauthor{saner2017masud} \cite{saner2015masud,saner2017masud} simply use TextRank on \emph{change request texts} for suggesting initial queries for concept location. \citet{dblp} build a term augmented tuple graph and use a random walk approach to reformulate queries for structured bibliographic DBLP Data (i.e., non-source code).  Ours is significantly different from these studies in the sense that we reformulate the initial queries not only by employing our term weighting method--CodeRank for \emph{source code}, but also by applying source code document structures, query quality analysis and machine learning. 
Besides, their reported best performance (\ie\ 58\%--62\% query improvement over baseline \cite{saner2017masud}) is quite lower than 
our performance (\ie\ 71\%, even with difficult queries). Given that reformulation is often performed on the initial queries, our technique can potentially complement theirs.
%There exist other studies \cite{ccmapping,infer,ase2016masud} that are also relevant to ours.
\citet{ccmapping} map method signatures to associated comments for query reformulation, and thus, might not work well with source code without comments.
\citet{ase2016masud} exploit crowd sourced knowledge for query reformulation, and their method is also subject to the availability of a third party information source.
Thus, while earlier studies adopt various methodologies or information sources, our technique not only employs a novel and promising term weight --\emph{CodeRank}, but also
exploits structures of the source documents for identifying the best reformulation to a given query for improved concept location.
%(using query quality analysis and machine learning) for improved query reformulation.

%\vspace{-.2cm}
\section{Conclusion \& Future Work}
\label{sec:conclusion}
To summarize, we propose a novel technique--ACER--for improved query reformulation for concept location.
It takes an initial query as input, identifies appropriate search terms from the source code using a novel term weight, and then suggests the best reformulation to the initial query using document structures, query quality analysis and machine learning.
% --CodeRank,
%and then suggests effective reformulation to the initial query by exploiting the source document structures, query quality analysis and machine learning.
%We employ a novel term weighting method--CodeRank-- for extracting candidate reformulations from the source code, and then exploit source document structures to suggest the best reformulation for a given query.
Experiments with 1,675 baseline queries from eight systems report that our technique can improve 71\% of the baseline queries and preserve 26\% of them, which are highly promising.
%349 change tasks from two  subject systems show that our approach, on average, can return relevant results (\ie\ Java classes) for 49.34\% of the change tasks with 57.16\% \emph{mean average precision} and 63.33\% \emph{recall}.
Comparison with five closely related approaches including the state-of-the-art not only validates our empirical findings but also demonstrates the high potential of our technique.
In future, we plan to apply our term weighting method, CodeRank, to other SE text retrieval tasks involving source code such as bug localization and traceability recovery.

% one of the latest and closely related state-of-the-art approaches also shows that our approach performs comparatively better in different performance metrics. 
%While these preliminary findings demonstrate the high potential of the proposed approach, further validation with more subject systems of diverse varieties and change tasks is warranted. 
%Given all these preliminary findings that demonstrate the potential of our technique, it must be further validated with more subject systems and more change tasks.

\section*{Acknowledgement}
This research work was supported in part by the Natural Sciences and Engineering Research Council of Canada (NSERC) and the Internal Dean's Scholarship of University of Saskatchewan, Canada.

\balance

%However, the finding and the model must be evaluated with larger dataset to reach a reliable level.
%\vspace{-.3cm}

\bibliographystyle{plainnat}
\setlength{\bibsep}{0pt plus 0.3ex}
%\scriptsize
\bibliography{sigproc}  % sigproc.bib is the name of the Bibliography in this case
% ACM needs 'a single self-contained file'!
%\balancecolumns
\end{document}